\magnification1200

\tolerance=2400

\baselineskip14pt

 at 10truept

\font\reg=cmr10 at 12truept

\font\large= cmr16 at 16truept

\font\llarge=cmr20 at 20truept

\font\lllarge=cmr24 at 24truept

\lllarge

\centerline{ Constraints on Automorphic Forms of Higher} \bigskip \centerline{Derivative Terms from
Compactification}

\vskip 2cm

\large
\centerline{Finn Gubay, Neil Lambert   and Peter West}

\bigskip

\centerline{Department of Mathematics}

\centerline{King's College London}

\centerline{The Strand, London}

\centerline{ WC2R 2LS, UK}

\bigskip
\reg \centerline{finn.gubay@kcl.ac.uk,\ neil.lambert@kcl.ac.uk,\
peter.west@kcl.ac.uk}

\bigskip

\llarge \centerline{\sl Abstract}

\medskip
\reg
 By dimensionally reducing the higher derivative corrections of
ten-dimensional IIB theory on a torus we deduce constraints on the
$E_{n+1}$ automorphic forms that occur in $d=10-n$ dimensions. In
particular we argue that these automorphic forms involve the
representation of $E_{n+1}$ with fundamental weight
$\vec\lambda^{n+1}$, which is also the representation to which the
string charges in $d$ dimensions belong. We also consider a similar
calculation for the reduction of higher derivative terms in
eleven-dimensional M-theory.

\bigskip \noindent

\vfill \eject

{\llarge {1. Introduction}}
\bigskip

It is well-known that the type IIA [1-3]  and type IIB [4-6] supergravity theories in $10$-dimensions are the
low energy effective actions (at second order in derivatives) of the type II string theories. Furthermore
Eleven-dimensional supergravity [7] is the low energy effective action of one of the limits of  M-theory. Upon
dimensional reduction on a torus to $d=10-n$ dimensions all these theories become equivalent and furthermore
posses a large and unexpected duality symmetry $E_{n+1}$ [8-11]

These supergravity theories are important as they encode all
perturbative and non-pertubative effects, many of which cannot yet
be calculated from first principles in String Theory or M-theory.
The higher derivative terms of the effective action also encode much
more structure of the fundamental theory and their study is of
considerable interest. Since brane charges must be quantised [12,13]
and are rotated into each other under $E_{n+1}$, only a discrete
subgroup of $E_{n+1}$ can be preserved in the quantum theory. This
led to conjectures of duality symmetries in four-dimensional String
Theory [14,15] and then  unified into so-called U-duality [16] for
all dimensions.

The study of U-duality groups and higher derivative terms was begun in [17] for the $R^4$ term of
ten-dimensional type IIB string theory with U-duality group $SL(2,{\bf Z})$   and has been considerably
extended to other terms [18-23]. Results also exist for higher derivative terms in less than ten dimensions
[23-36]. The effective action consists of polynomial in the field strengths, Riemann curvatures, and
derivatives of the scalar fields multiplied by functions of the scalar fields. These functions
%, which are not analytic functions of the scalars,
transform in a simple way under discrete $E_{n+1}$ transformations
and can be identified with non-holomorphic automorphic forms. In
fact in most terms considered so far the functions transform
trivially.

The theory of automorphic forms is complicated and still emerging. Large classes of these objects can be
constructed by specifying a particular representation of a group $G$ along with a suitable subgroup $H$. In
String Theory $G$ can be identified with the duality group $E_{n+1}$ and $H=I(E_{n+1})$ with  Cartan involution
invariant subgroup (which is also the maximally compact subgroup) and the scalar fields are known to
parameterise an $E_{n+1}/I(E_{n+1})$ coset.

An important problem in String Theory is to identify the relevant
automorphic forms that arise in various dimensions as coefficients
of the higher derivative terms. One might  hope that there are some
relatively simple organising principles that control which
automorphic forms appear. One of which is supersymmetry  which
relates   various higher derivative terms  of the same order in
spacetime derivatives. Once a suitable automorphic form is known one
can then in principle read off  its perturbative and
non-perturbative parts and hence deduce important information about
the underlying quantum theory, such as non-renormalisation and
instanton effects. These can then be checked against String Theory
calculations  [17-36] and non-renormalisation theorems [37].

One important  question is:  which representations enter into the
construction of the automorphic forms? We will investigate this
problem in this paper  by examining the $E_{n+1}$ group structure of
higher derivative terms in type IIB String theory and M-theory,
after they have been dimensionally reduced to $d$ dimensions. We
will see that this will enable us to deduce constraints on the
representations that appear for general higher derivative terms in
$d$-dimensions. Our results are very general and are consistent with
conjectures   and other results for special cases [27,35,36].

This paper is a continuation of the analysis found in [28,29]. In particular we will dimensionally reduce a
generic higher derivative term of type IIB String Theory  on an $n$- torus to $d=10-n$ dimensions. When one
does this one finds that each  term in the $d$ dimensional effective action contains a factor  of the form
$e^{{ \sqrt{2}}\vec w\cdot\vec\phi}$ for some vector $\vec w$. The fields $\vec \phi$ are  the diagonal
components of the metric, which encode the volume and other radii moduli of the torus, as well as any scalar
fields in the higher dimensional theory such as  the dilaton. Carrying out this for the supergravity theory,
that is the lowest energy terms, we find that the vectors $\vec w$ that appear are the roots of $E_{n+1}$.
Indeed, this provides the fastest way to see that there is very likely to be an $E_{n+1}$ symmetry of the lower
dimensional theory. In references  [28,29] this was carried out for a generic higher derivative term of the
effective action and one found not roots but weights of $E_{n+1}$. This in itself was evidence for an $E_{n+1}$
symmetry in the higher derivative corrections to string theory and the appearance of weights rather than roots
was interpreted as evidence for automorphic forms as they involve weights of $E_{n+1}$. In this paper we take
this work to its natural conclusion and compute the weights that should appear in the automorphic forms.

We carry out the calculation in a more streamlined manner than in references [28,29] and generalise to any
dimension $d\ge 3$. We assemble the fields of the lower dimensional theory, that occur with spacetime
derivatives, into representations of the $d$-dimensional duality group $E_{n+1}$. We show that the higher
derivative  terms can be written as powers of the $E_{n+1}$ covariant field strengths along with additional
factors of the dilaton and volume which are the remnants of the above $e^{{\sqrt{2}}\vec w\cdot\vec\phi}$
factors. We find that the left-over weight has a simple  universal form for any term. For terms that arise at
tree level in string perturbation in ten dimensions we find $\vec w= s\vec\lambda^{n+1}$, where $s=(l_T-2)/4$
with  $l_T$ counting the number of derivatives and $\vec\lambda^{n+1}$ the fundamental weight dual to
$\vec\alpha_{n+1}$ (see Figure 1). The observation of [29] is that these additional factors must come from an
automorphic form and therefore we are led to conclude that the automorphic form which multiplies a given higher
derivative term involves the weight $\vec\lambda^{n+1}$. Moreover, for Eisenstein-like automorphic forms the
leading order behaviour is given by $e^{-{\sqrt{2}}s\vec \phi\cdot\vec\lambda^H}$ ( for example see [29,30]),
where $\vec\lambda^H$ is the highest weight of the representation used to construct the automorphic form. Thus
our results suggest that the higher derivative terms always include an automorphic form built from a
representation with  highest weight $\vec \lambda^{n+1}$. We also perform a similar  analysis in M-theory  and
see that the weight is (using the same labeling of the $E_{n+1}$ diagram), $\vec w = s\vec\lambda^{n-1}$ with
$s=(l_T-2)/6$. However it is important to note that we are in effect considering a particular limit and other
representations could also appear but be subdominant in that limit.

The rest of this paper is organised as follows.  In section 2 we perform our dimensional reduction analysis for
type IIB string theory. In particular we perform a dimensional reduction with manifest $SL(2)\times SL(n)$
symmetry and show how this can be embedded into an $E_{n+1}$ symmetry. We include a detailed discussion of our
methods as well as a description of $SL(2)\times SL(n)$ and $E_{n+1}$ coset constructions, including explicit
examples. In section 3 we perform a similar analysis for the reduction of higher derivative terms of M-theory
with a  manifest $SL(n)$ symmetry. As mentioned above this leads to different weights of $E_{n+1}$ to that
obtained in type IIB. In section 4 we conclude with a discussion of our results. We also include appendix A
with a short review on the coset construction.

\bigskip

{\llarge 2. Dimensional reduction of type IIB higher derivative
 terms}
\bigskip

In this section we will study the dimensional reduction of the
higher derivative  corrections of IIB string theory. Our methods
follow those of section two of reference [28],  suitably generalised
for the reduction on an $n$ torus to any dimension rather than the
three dimensions considered there. Our  metric compactification
ansatz  is given by
$$
d\hat s^2 = e^{2\alpha \rho}ds^2 + e^{2\beta\rho}G_{ij} (dx^i + A^i_\mu dx^\mu) (dx^j + A^j_\mu dx^\mu)\ ,
\eqno(2.1)
$$
where $G_{ij}$ has unit determinant and
$$
\alpha  = \sqrt{{n}\over 2(D-2)(d-2)}\ ,\qquad \beta = -{(d-2)\alpha \over n}\ . \eqno(2.2)
$$
These values of $\alpha$ and $\beta$ ensure that, provided one starts in $D$-dimensional Einstein frame, the
resulting theory in $d$ dimensions is in Einstein frame with a standard normalisation for the kinetic term of
the scalar $\rho$ which controls the volume of the torus. We have labeled the coordinates by $(x^\mu , x^i),
\mu=0,1,\ldots ,d-1;\ i=d, \ldots , D-1$. In the above equation $G_{ij} = e_i^{\ \overline k}e_j^{\ \overline
l}\delta_{\bar k\bar l}$ and  $e_i^{\ \overline k}$ is a vielbein but subject to  $\det e =1$. We adopt the
convention that $i,j,k,\ldots $ are world indices and $\overline i,\overline j,\overline k,\ldots $ are tangent
indices.

As is well-known, dimensionally reducing Einstein gravity on a torus leads to a theory that possesses an
$SL(n,{\bf R})$ symmetry. In particular, the degrees of freedom of gravity associated with the torus, apart
from the graviphotons, enter the lower dimensional theory through a non-linear realization of $SL(n,{\bf R})$
with local subgroup $SO(n)$. The latter is the Cartan involution invariant subgroup, {\it i.e.}
$I(SL(n))=SO(n)$. This is to be expected as in the theory before the dimensional reduction the gravity degrees
of freedom are in the non-linear realisation of $ISL(D)$ with local subgroup $SO(D)$.

The non-linear realisation is discussed for an arbitrary group in Appendix A. Using the local subgroup we can
bring the $SL(n)$ group element to the form
$$
g_{sl_n}(\xi_{sl_n}) = e^{\sum_{\underline \alpha
 >0}E_{\underline\alpha} \chi_{\underline\alpha} }e^{-{{1}\over {\sqrt
{2}}}\underline\phi\cdot \underline H} \eqno(2.3)
$$
where $\underline H$ forms the Cartan subalgebra, $E_{\underline
\alpha}$ are positive root generators (when $\underline\alpha>0$) of
$SL(n, {\bf R})$ respectively and $\xi_{sl_n}$ collectively denotes
the fields $ \chi_{\underline\alpha} $ and $\underline\phi $.  In
fact the terms which contain $g_{sl_n}(\xi_{sl_n})$ alone are  built
out of the Cartan forms
$$
g_{sl_n}^{-1}\partial_\mu g_{sl_n} =P_{sl_n\mu}+Q_{sl_n\mu}\ ,
\eqno(2.4)
$$
where $P_{{sl_n}\mu}$ and $Q_{{sl_n}\mu}$ are symmetric and anti-symmetric in $\overline i$ and $\overline j$
respectively corresponding to the decomposition of the Cartan forms into those for $SO(n)$, {\it i.e.}
$Q_{{sl_n}\mu}$, and its compliment.

In what follows we will construct the dimensionally reduced theory
in such a way that its $SL(n,{\bf R})$ symmetry is manifest. To
begin with we wish to find an expression for the inverse vielbein
making use of the discussion  of non-linear realisations (see
appendix A). Let us denote the  the fundamental highest weights of
$SL(n)$ by $\underline \lambda^i$.   The representation with highest
weight $\underline \lambda^1$  corresponds to the vector
representation, with a single lowered index. We denote the states of
this representation by $|\psi>=\psi_{i }|\underline \mu^{i}>$ where
$\underline\mu^i$ are the weights in the root string of
$\underline\lambda^1$, which we denote by $[\underline\lambda^1]$.
From this linear representation we can construct the non-linearly
transforming representation using equation (A.7) as follows
$$
|\varphi(\xi)> = \sum \varphi_i |\vec\mu^i> =L(g^{-1}_{sl_n}(\xi))|\psi> = e^{{1\over\sqrt{2}}\vec\phi\cdot\vec
H} e^{\sum_{\vec\alpha>0} -\chi_{\vec\alpha} E_{\vec\alpha} }|\psi> \ ,\eqno(2.5)
$$
and so we write
$$
\varphi_i=D(g_{sl_n}^{-1}(\xi_{sl_n}))_{i}^{\ j}\psi_j\ . \eqno(2.6)
$$
Under an $SL(n)$ transformation this state  transforms under a local $SO(n)$ and we may identify the change
from $\psi_i$ to $\varphi_i$ as the familiar conversion from world to tangent indices using the inverse
vielbein. The matrix element of $g^{-1}_{sl_n}$ in the vector representation is therefore given by
$$
(e^{-1})_{\overline i}{}^j= D(g_{sl_n}^{-1}(\xi_{sl_n}))_{i}^{\ j}\ . \eqno(2.7)
$$
The right-hand end of equation (2.5) contains the factor $e^{{{1}\over {\sqrt{2}}}\underline\phi\cdot
[\underline \lambda^{1}]}$. Thus we find that the inverse vielbein $e_{ i}{}^{\overline j}$ contains factors of
$e^{-{{1}\over {\sqrt{2}}}\underline \phi\cdot [\underline \lambda^1]}$.

The dimensionally reduced theory will involve corrections that
contain field strengths of the form ${\cal F}_{\mu_1\ldots \mu_p
i_1\ldots i_k}$, where $i_1, \ldots i_k$ are worldvolume indices of
the torus (there can also be $SL(2)$ indices which we address
below). However, we can choose to work with tangent, rather than
world, indices in the torus directions by using  the inverse
vielbein $(e^{-1})_{\overline i}^{\ j}$. Following the same argument
we just used above, this can be viewed as the conversion of the
linear rank $k$ antisymmetric representation of $SL(n,{\bf R})$ into
a non-linear representation of $SL(n,{\bf R})/SO(n)$ whose indices
rotate under $SO(n)$. Thus we start from the linear representation
$\sum_{i_1,\ldots i_k} {F}_{\mu_1\ldots \mu_p i_1\ldots i_k} |
i_1\ldots i_k,\underline \lambda^k >$ and construct the non-linear
realisation
$$
\sum_{i_1,\ldots i_k} {\cal F}^{sl(n)}_{\mu_1\ldots \mu_p \overline i_1\ldots \overline i_k} | i_1\ldots
i_k,\underline \lambda^k
>= L(g^{-1}_{sl_n}(\xi)) \sum_{i_1,\ldots i_k} F_{\mu_1\ldots \mu_p  i_1\ldots i_k} | i_1\ldots i_k,\underline \lambda^k
> \ .\eqno(2.8)
$$
We note that we have denoted the field strength with tangent indices
by ${\cal F}^{sl(n)}_{\mu_1\ldots \mu_p \overline i_1\ldots\overline
i_k}$. Its  dependence on the metric of the torus is obtained by
acting with $L(g^{-1}_{sl_n}(\xi))$ on the states $| i_1\ldots
i_k,\underline\lambda^k>$. Therefore one finds that the fields
$\underline \phi$ associated with the Cartan subalgebra of
$SL(n,{\bf R})$ occur in ${\cal F}^{sl(n)}_{\mu_1\ldots \mu_p
\overline i_1\ldots\overline i_k}$ through the factor $e^{{1\over
\sqrt{2}}\underline\phi\cdot [\underline \lambda^{k}]}$.

\bigskip{\bf{ 2.1 Review of $SL(2)$ Formulation of type IIB Supergravity}}\bigskip

We can treat the $SL(2, {\bf R})$ indices that arise in type IIB
supergravity in a similar way. To illustrate this let us review in
detail the $SL(2)$ invariant form of the ten-dimensional type IIB
supergravity.

In ten dimensions the scalars belong to the non-linear realisation
of $SL(2, {\bf R})$ with local subgroup $SO(2)$. The $SL(2, {\bf
R})$ group element can be brought to the form
$$
g_{sl_2}(\tau) = e^{E \chi }e^{-{1\over \sqrt{2}}\phi  H}\ , \eqno(2.9)
$$
where $E$ and $H$ are the  positive root and Cartan subalgebra
generators of $SL(2,{\bf R})$ respectively.   The scalars appear
through the $SL(2, {\bf R})$ Cartan form
$$
g_{sl_2}^{-1} \partial_\mu g_{sl_2}(\tau)=P_\mu^{sl_2}+Q_\mu^{sl_2}= -{1\over \sqrt 2} \partial_\mu\phi
H+\partial_\mu \chi e^\phi E .\eqno(2.10)
$$
Thus
$$\eqalign{
P^{sl_2}_\mu&=-{1\over \sqrt 2}\partial_\mu \phi H+\partial_\mu \chi
e^\phi {(E+F)\over 2} \equiv  P_{\mu 1}{H\over\sqrt{2}}+ P_{\mu 2}
{(E+F)\over 2},\cr Q^{sl_2}_\mu &= \partial_\mu \chi e^\phi
{(E-F)\over 2}\ .} \eqno(2.11)
$$
Under a local transformation $h=exp(\theta {(E-F)\over 2})$, the
Cartan forms transform as $P_\mu\to hP_\mu h^{-1}$ and so $P_{\mu
1}\to cos\theta P_{\mu 1}+sin\theta P_{\mu 2}$, $P_{\mu 2}\to
cos\theta P_{\mu 2}-sin\theta P_{\mu 1}$. As a result we find that
the complex quantity $P^+_{\mu }\equiv (P_{\mu 1}+ iP_{\mu 2})=
-ie^{\phi}\partial _\mu ( -\chi+ie^{-\phi})$ transforms as
$P^+_{\mu}\to e^{ i\theta}P^+_{\mu}$.  An invariant action for the
scalars is then given by
$$\eqalign{
S_{scalar}&=- {1\over 2}\int d^{10} x\ {\rm det}e P_{\mu }^-P^{+\mu} \cr &= - {1\over 2}\int d^{10}x \ {\rm
det} e(\partial_\mu \phi
\partial^\mu \phi+\partial_\mu\chi\partial^\mu\chi
e^{2\phi})\cr &= - {1\over 2}\int d^{10} x \ {\rm det}e {|\partial_\mu\tau |^2\over (Im\tau)^2}\ , }\eqno(2.12)
$$
where $P_\mu^- = (P_\mu^+)^*$ and in the last line we have
introduced $\tau=-\chi+ie^{-\phi}$; $\tau$ undergoes fractional
linear transformations under the action of $SL(2,{\bf R})$ on this
coset.

The IIB theory in ten dimensions contains a five-form field strength
$F_{\mu_1\ldots \mu_5}$, which is a singlet under $SL(2, {\bf R})$.
There are also two three-form field strengths $F_{\mu_1\mu_2\mu_3
\alpha}=3\partial _{[\mu_1}A_{\mu_2\mu_3]\alpha}$, $\alpha=1,2$
which transform as the linear doublet representation of $SL(2,{\bf
R})$. We can form the non-linear realisation by acting with
$L(g_{sl_2}^{-1})$, on this representation; we find, using equation
(A.7),  the corresponding non-linear representation;
$$
{\cal  F}^{sl_2}_{\mu_1\mu_2 \mu_3 \alpha} |\alpha , \mu >= L(g^{-1}_{sl_2}(\tau))F_{\mu_1\ldots \mu_3 \alpha
}|\alpha , \mu
>\ , \eqno(2.13)
$$
where $ |1 , \mu >=|\mu >$ and  $ |2 , \mu >=|-\mu
>$ with $\mu={1\over \sqrt 2}$ being the fundamental representation
of $SL(2, {\bf R})$. Viewing ${\cal
F}^{sl_2}_{\mu_1\mu_2\mu_3\alpha}$ as a column vector we find that
$$
{\cal F}^{sl_2}_{ \mu_1\mu_2 \mu_3 }= \left(\matrix {e^{{\phi\over
2}}( F_{1 \mu\nu\rho}-\chi F_{2 \mu\nu\rho} )\cr e^{-{\phi\over 2}}
F_{2 \mu\nu\rho}\cr }\right)\ , \eqno(2.14)
$$
which now only transform sunder the local subgroup $SO(2)$. Introducing the complex quantity ${\cal  F}_{
\mu\nu\rho}^+$ this transformation becomes
$$
{\cal F}_{ \mu\nu\rho}^+\equiv {\cal  F}_{1 \mu\nu\rho}+ i {\cal F}_{2 \mu\nu\rho}= e^{{\phi\over 2}}(F_{1
\mu\nu\rho }+\tau F_{2 \mu\nu\rho}) \to e^{ i\theta } {\cal F}_{ \mu\nu\rho} ^+\ . \eqno(2.15)
$$
A manifestly  $SL(2,{\bf R})$ invariant action is given by
$$
S_{2form} = \int d^{10} x\  {\rm det}e {\cal F}_{ \mu_1\mu_2\mu_3}^+ {\cal F}_{ \nu_1\nu_2\nu_3}^- =\int d^{10}
x \ {\rm det} e(e^\phi (F_{1}-\chi F_{2} )^2+e^{-\phi}F_{2}^2)\ , \eqno(2.16)
$$
where $ {\cal F}_{ \mu_1\mu_2\mu_3}^-=( F_{ \mu_1\mu_2\mu_3}^+)^*$.

Including the standard Einstein-Hilbert term and a kinetic term for
the 5-form field strength we arrive at the Bosonic part of the type
IIB action
$$
S_{IIB} =\int d^{10} x\  {\rm det}e (R -P_{\mu }^+P^{\mu -}- {1\over
2\cdot 3!}{\cal F}_{ \mu_1\mu_2\mu_3}^+ {\cal F}^{
\mu_1\mu_2\mu_3-}-{1\over 2\cdot 5!}\tilde F_{\mu_1\ldots\mu_5}
\tilde F^{\mu_1\ldots\mu_5} )\ .\eqno(2.17)
$$
Note that the five-form field strength equation of motion must be supplemented by  a self-duality constraint
and $\tilde F_{\mu_1\ldots\mu_5}=F_{\mu_1\ldots\mu_5}$ up to terms involving the 2-forms. Making the transition
to string frame $e_\mu{}^{\overline \nu}=e^{-{\phi\over 4}}(e_S)_\mu{}^{\overline \nu}$, the type IIB effective
action becomes
$$\eqalign{
S_{IIB}&=\int d^{10} x \ {\rm det}(e_S)  (e^{-2\phi}R-{1\over 2}e^{-
2\phi}\partial_\mu \phi
\partial^\mu \phi-{1\over 2}\partial_\mu\chi\partial^\mu\chi\cr
&\hskip3cm -{1\over 2\cdot3!}  (F_{1}-\chi F_{2} )^2-{1\over
2.3!}e^{-2\phi}F_{2}^2-{1\over 2\cdot 5!}\tilde
F_{\mu_1\ldots\mu_5}\tilde F^{\mu_1\ldots\mu_5})\ .} \eqno(2.18)
$$
We note from the factors of $e^\phi $  that occur that
$F_{1\mu_1\mu_2\mu_3}$, $P_{\mu2}$, $\chi$ and $F_{\mu_1\ldots
\mu_5}$ are in the R-R sector and $g_{\mu\nu}$, $\phi$, $P_{\mu1}$
and $F_{2\mu_1\mu_2\mu_3}$ in the NS-NS sector.

Let us now consider higher derivative terms. It will be useful for
what follows to use a hat to denote a ten-dimensional quantity and
suppress any spacetime indices. The higher derivative corrections in
the IIB theory in ten dimensions can be written as a  polynomial in
the Riemann tensor $\hat R$, $\hat P^{sl_2}$,  rank three field
strength ${\cal F}_{3}^{\pm}$ and  rank five field strength $F_{5}$
with coefficients that are $SL(2)$ automorphic forms. The generic
term has the form
$$
\int d^{10} x \ {\rm det}( \hat e)\partial^{\hat l_0} \hat R^{{\hat l_R\over 2}}(\hat P^{sl_2}_{\mu_11})^{\hat
l_1} (\hat P^{sl_2}_{\mu_12})^{\hat l'_1}( \hat {\cal F}_{2\mu_1\ldots\mu_3})^{\hat l_3}( \hat {\cal
F}_{1\mu_1\ldots \mu_3 })^{\hat l_3'}(\hat  F_{\mu_1\ldots \mu_5})^{\hat l'_5} \hat\Phi_{sl_2} \eqno(2.19)
$$
where $\hat\Phi_{sl_2}$ is a suitable automorphic form. As is well know the higher order corrections involve
instantons and other solitonic objects and due to the quantisation conditions on the charges only the $SL(2,Z)$
part of the $SL(2,R)$ symmetry survives. The automorphic form depends on $\tau$ that is $\phi$ and $\chi$. We
will be mainly interested in the $e^\phi$ dependence and we denote the leading dependence of $\hat \Phi_{sl2}$
on $\phi$ by $\hat \Phi_{sl2}\sim e^{-{\hat s} \phi}$.

It will be instructive to compute the $e^\phi$ dependence of the above ten dimensional higher derivative
correction in string frame. The transition from Einstein frame to  string frame is given by $\hat e=
e^{-{\phi\over 4}}\hat e_s$.  We find that the above term leads to the factor
$$
e^{{\phi\over 4}(\hat l_0+\hat l_R+\hat l_1+5\hat l'_1+\hat l_3 + 5\hat l_3'+5\hat l'_5-10-4\hat s)}\ .
\eqno(2.20)
$$
Note that we have used a prime to denote contributions from R-R fields. At order $g$ in perturbation theory we
have the contribution $e^{\phi(2g-2)}$ and so we conclude that for a perturbative contribution
$$
\hat s= {1\over 4}(\hat l_0+\hat l_R+\hat l_1+5\hat l'_1 +\hat l_3+5\hat l'_3 + 5\hat l'_5-2-8g)\ . \eqno(2.21)
$$

\bigskip{\bf{ 2.2 Reduction of Higher Derivative Type IIB Terms}}\bigskip

In this paper we are interested in the dimensional reduction of ten
dimensional higher derivative corrections of IIB string theory, that
is terms as given in equation (2.19), on an $n$ torus to $d=10-n$
dimensions.  As explained above, by working with the non-linear
realisations we can formulate the result with a  manifestly
$SL(2)\otimes SL(n)$ symmetry. The resulting building blocks in $d$
dimensions are  the Riemann tensor $R$ which is a $SL(2)\otimes
SL(n)$ singlet, the derivatives of the  scalars which belong to the
Cartan forms of $SL(2)\otimes SL(n)$, $P_{sl_2 \otimes sl_n}$ and
objects which are non-linear representations of $SL(2)\otimes
SL(n)$. As mentioned above the latter arise if one  works with
``tangent space" quantities. These objects generically denoted by
${\cal F}$ are related to the usual  field strengths, which
transform linearly under $SL(2)\otimes SL(n)$, to the non-linearly
realised objects $F $,  by the generic equation
$$
|{\cal F}_{sl_2 \otimes sl_n}>= L(g^{-1}_{sl_2 \otimes sl_n}) |F >
\equiv e^{{{1}\over {\sqrt {2}}}(\underline\phi\cdot \underline
H+\phi  H)} e^{-(\sum_{\underline \alpha
 >0}E_{\underline\alpha} \chi_{\underline\alpha} +E \chi)}|F >\ .
\eqno(2.22)
$$
These $\cal F$ transform by  field dependent $SO(n)\otimes SO(2)$
transformations and so it is easy to construct invariants using the
Kronecker delta symbol.  The $Q_{sl_2 \otimes sl_n}$ component of
the Cartan forms only enters when we find derivatives of the above
objects where it plays the role of a connection.

We are particularly interested of the dependence in the dimensionally reduced action on the scalars $
\phi,\rho$ and $\underline \phi$ which we assemble into the $n+1$-vector
$$
\vec \phi = (\phi,\rho,\underline\phi)\ . \eqno(2.23)
$$
The dependence on $\phi$ and $\underline \phi$, which are the Cartan fields associated with $SL(2)\otimes
SL(n)$, occurs only inside the objects ${\cal F}_{sl_2 \otimes sl_n}$. The exception is the $\phi$ dependence
that arises from the ten dimensional automorphic form $\hat \Phi_{sl_2}$. The dependence on  $\rho$ arises from
the dimensional reduction of the   vielbeins using  the metric ansatz of equation (2.1) as was described in
references [28,29]. The $\det\hat e$ factor in the action leads to a factor of
$e^{(d\alpha+n\beta)\rho}=e^{2\alpha\rho} $ while ${\cal F}_{sl_2 \otimes sl_n\  \mu_1\ldots \mu_p \overline
i_1\ldots\overline i_k}$ leads to the factor $e^{-\rho(p\alpha+k\beta)}$. To give a concrete example with $l$
factors of the latter field strength we find the generic term
$$\int d^{10} x  \det \hat e (\hat {\cal F}^{sl_2}_{ \mu_1\ldots \mu_{q}})^l \sim \sum_{p+k=q} \int d^d x \det e ({\cal F}_{sl_2 \otimes sl_n\ \mu_1\ldots \mu_p
\overline i_1\ldots\overline i_k})^l
e^{(2\alpha-l(p\alpha+k\beta))\rho}\ . \eqno(2.24)
$$
The powers of
$e^\rho$  associated with any other terms are also easily
calculated.

The dimensional reduction of  any term in the effective action of
equation (2.17) leads to terms that contain the derivative of
scalars, vierbein and gauge fields  multiplied by factors of the
form $e^{\sqrt{2}\vec w\cdot\vec \phi}$ for some $n+1$-vector $\vec
w$:
$$
\vec w = (w,\kappa,\underline w) \ .\eqno(2.25)
$$
The first and third entries  $w$ and $\underline w$ arise  from the
behaviour of the fields under the $SL(2,{\bf R})\otimes SL(n)$ and
can be read off from the action of $g_{sl_2 \otimes sl_n}$ on the
linearly realised representation using equation (2.22). The second
entry simply records the powers of $e^{\sqrt{2}\rho}$ that arise
after dimensional reduction as just discussed.

For every factor of ${F}_{ \mu_1\mu_2\mu_3}^{sl_2}$  that occurs one
finds a corresponding factor of $e^{{{1}\over {\sqrt{2}}}\phi
[\mu]}$, where $[\mu] =\{{1\over \sqrt{2}},-{1\over\sqrt{2}}\} $ are
the weights that appear in the fundamental representation of
$SL(2,{\bf R})$. In particular the NS-NS and R-R field strengths
come with the factors $e^{-{\phi\over 2}}$ and $e^{{\phi\over 2}}$
respectively as are most easily read off from equation (2.14).

In what follows it will be advantageous to also consider the dual
version of certain fields.  Let us consider a two-derivative term in
the low energy effective action of the form
$$
\int d^d x \det e ({\cal F}_{sl_2 \otimes sl_n \mu_1\ldots \mu_p \overline i_1\ldots\overline i_k})^2
e^{(2\alpha-2(p\alpha+k\beta)\rho} \ ,\eqno(2.26)
$$
where ${\cal F}_{sl_2 \otimes sl_n} = g^{-1}_{sl_2\otimes sl_n}F$, $F=dA$. We can introduce the dual field
strength ${\cal F}^D_{sl_2 \otimes sl_n \mu_1\ldots \mu_q \overline i_1\ldots\overline i_s}$ defined by ${\cal
F}^D_{sl_2 \otimes sl_n}= g^{-1}_{sl_2\otimes sl_n}dA^D$ where $p+q=d$ and $k+s=n$. We then impose the Bianchi
identity of $F=dA$ by adding to the action the term
$$\eqalign{
\int d^d x &\epsilon^{\mu_1\ldots \mu_p \nu_1\ldots
\nu_q}\epsilon^{i_1\ldots\overline i_k j_1\ldots\overline i_s}{
F}_{sl_2 \otimes sl_n \mu_1\ldots \mu_p \overline i_1\ldots\overline
i_k} {F}^D_{sl_2 \otimes sl_n \nu_1\ldots \nu_q \overline
j_1\ldots\overline j_s}\cr & = \int d^d x \epsilon^{\mu_1\ldots
\mu_p \nu_1\ldots \nu_q}\epsilon^{i_1\ldots\overline i_k
j_1\ldots\overline i_s}{\cal F}_{sl_2 \otimes sl_n \mu_1\ldots \mu_p
\overline i_1\ldots\overline i_k} {\cal F}^D_{sl_2 \otimes sl_n
\nu_1\ldots \nu_q \overline j_1\ldots\overline j_s}\ ,} \eqno(2.27)
$$
where $F^D = dA^D$ and in the second line with have used the fact
that $\det(g_{sl_2\otimes sl_n})=1$. Note that if ${\cal F}_{sl_2
\otimes sl_n}$ and ${\cal F}^D_{sl_2 \otimes sl_n}$ have  $SL(2)$
indices then an additional factor of $\epsilon^{ab}$ is needed in
(2.27).

We can now view ${\cal F}_{sl_2 \otimes sl_n}$ as an unconstrained field and integrate it out. Taking its
equation of  motion implies that
$$
{\cal F}_{sl_2 \otimes sl_n \mu_1\ldots \mu_p \overline
i_1\ldots\overline i_k}\sim \epsilon^{i_1\ldots\overline i_k
j_1\ldots\overline i_s} \epsilon^{\mu_1\ldots \mu_p \nu_1\ldots
\nu_q} {\cal F}^D_{sl_2 \otimes sl_n \nu_1\ldots \nu_q \overline
j_1\ldots\overline j_s} e^{2(-\alpha+(p\alpha+k\beta))\rho}
\eqno(2.28)
$$

We will assume that we can use this lowest order dualisation
equation in the higher order corrections. Therefore,  for each
factor of ${\cal F}_{sl_2 \otimes sl_n\mu_1\ldots \mu_p \overline
i_1\ldots\overline i_k}$ we find in the higher derivative terms
$$\eqalign{
{\cal F}_{sl_2 \otimes sl_n \mu_1\ldots \mu_p \overline
i_1\ldots\overline i_k}e^{-(p\alpha+k\beta)\rho} &\sim {\cal
F}^D_{sl_2 \otimes sl_n \nu_1\ldots \nu_q \overline
j_1\ldots\overline j_s} e^{-2\alpha+(p\alpha+k\beta)\rho} \cr &\sim
{\cal F}^D_{sl_2 \otimes sl_n \nu_1\ldots \nu_q \overline
j_1\ldots\overline j_s} e^{-(q\alpha+s\beta)\rho} \ .}\eqno(2.29)
$$
In the last step used equation (2.2). Hence, we
get the same result if we use the original field or we use the dual field provided we  take into account the
correct number of indices. The reader may check this in specific cases including that of the graviphoton which
first appears when  reducing the Riemann tensor with a field strength that carries  a single upper $i$ index.

It is rather pleasing to compute the vectors $\vec w$ that arise when dimensionally reducing the IIB
supergravity theory of equation (2.17) and show that one finds the weights of $E_{n+1}$.

%%%%%%%%%%%%%%%%%%%%%%%%%%%%%%%%%%%%%%%%%%%%%%%%%%%%%%%%%%%%%%%

\bigskip
{\bf  2.3 The $E_{n+1}$ symmetry   in $d$ dimensions}
\bigskip

As discussed in the last section the dimensional reduction of the
IIB theory including its higher derivative corrections, on an $n$
torus leads to  a formulation in which the $SL(2)\otimes SL(n)$
symmetry is manifest. However, the IIB supergravity theory when
dimensionally reduced to $d=10-n$ dimensions   actually possess
an $E_{n+1}$ symmetry, of which a discrete subgroup is preserved in
the quantum theory. Evidence for this conjecture has been obtain in
a variety of works such as [17-36]. The Dynkin diagram of $E_{n+1}$
suited to the IIB theory is given by
$$
\matrix{ & & & & & & & &\bullet&\vec\alpha_{n+1} & \cr & & & & & & &
& |& & \cr & & & & & & & &\bullet&\vec\alpha_n& \cr & & & & & & &
&|& &\cr
\bullet&-&\ldots&-&\bullet&-&\bullet&-&\bullet&-&\bullet&\cr
\vec\alpha_{1}& && &\vec\alpha_{n-4}& &\vec\alpha_{n-3}& &
\vec\alpha_{n-2}& &\vec\alpha_{n-1}\cr}
$$
\bigskip
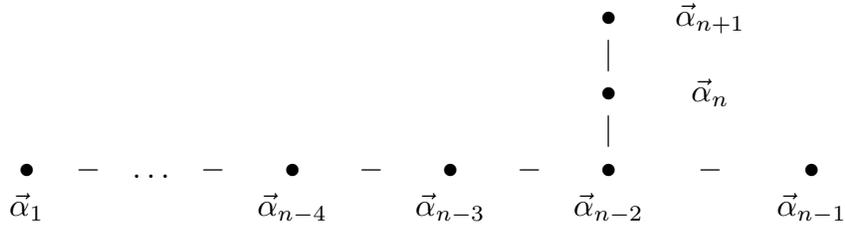
\centerline{ Fig 1: Dynkin diagram for $E_{n+1}$ in type IIB labelling}
\bigskip

The relevant $SL(2)\otimes SL(n)$ subalgebra of $E_{n+1}$ is found
by deleting the node labeled $n$ in the Dynkin diagram of Figure 1.
The $SL(2)$ factor is just the $SL(2)$ symmetry of the IIB
supergravity theory and arises from the node labeled $n+1$, while
the $SL(n)$ symmetry is part of the gravity symmetry of the ten
dimensional theory that now belongs to the torus and corresponds to
the nodes labeled 1 to $n-1$.

These features are particularly apparent when one considers the
$E_{11}$ formulation of the IIB theory [38,39]. The $E_{n+1}$ Dynkin
diagram emerges from the $E_{11}$ Dynkin diagram, given just below,
by deleting the node $d$ to find the algebra $SL(d)\otimes E_{n+1}$.

$$
\matrix{& & & & & & & &\bullet &\vec\alpha_{11} & \cr & & & & & & & &| & & \cr & & & & &&& &\bullet
&\vec\alpha_{10}& \cr & & & &&& & &| & & \cr \bullet&-&\bullet&-&\ldots &- &\bullet&-&\bullet&-&\bullet \cr
\vec\alpha_1& &\vec\alpha_2& & & &\vec\alpha_7& &\vec\alpha_8& & \vec\alpha_9\cr}
$$
\bigskip
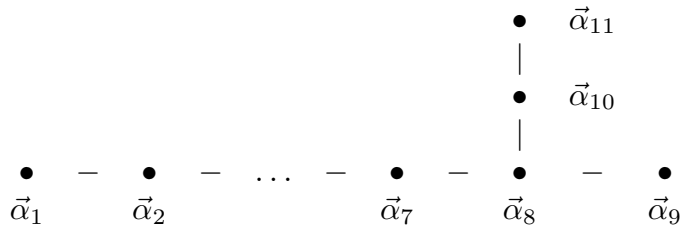
\centerline{Fig 2: Dynkin diagram for $E_{11}$}
\bigskip

The nodes labeled 1 to 9 of the $E_{11}$ Dynkin diagram are called the gravity line as they are associated with
ten dimensional gravity.  After the deletion of the node $d$, this line gives rise to $SL(d)\otimes SL(n)$
which is associated with gravity in $d$ dimensions and the $SL(n)$ of the now internal $E_{n+1}$ symmetry.

As already mentioned if  one computes the weights $\vec w$ that arise from the dimensional reduction of the IIB
supergravity theory using the techniques given in the last section one readily finds that they are  the weights
of $E_{n+1}$. While this is a strong indication of an underlying $E_{n+1}$ symmetry the detailed dimensional
reduction is required to prove the existence  of this symmetry in $d$ dimensions. In this process one finds
that the $SL(2)\otimes SL(n)$ representations that the fields belong to collect up to form a representation of
$E_{n+1}$. In this paper it will be essential to understand how the representations of $E_{n+1}$ that occur
decompose into representations of $SL(2)\otimes SL(n)$ as this will allow us to compare the $E_{n+1}$
formulation of the higher derivative corrections with that  arising from dimensional reduction from ten
dimensions. It is from this comparison that we will be able to deduce some properties of the automorphic form
in $d$ dimensions. The review [40] on U-duality discusses $E_{n+1}$ representations but here we will need the
explicit form for the weights.

\bigskip
 {\offinterlineskip \tabskip=0pt \halign{ \vrule height2.75ex depth1.25ex width 0.6pt #\tabskip=1em
& \hfil #\hfil &\vrule \hfil #\hfil & \hfil #\hfil &\vrule #  &
\hfil #\hfil &\vrule #  & \hfil #\hfil &\vrule #  &  \hfil
#\hfil&\vrule # &  \hfil #\hfil&\vrule # &\hfil #\hfil & #\vrule
width 0.6pt \tabskip=0pt\cr \noalign{\hrule height 0.6pt} & \omit
$d$ &&\omit $E_{n+1}$ && \omit $I(E_{n+1})$ &&\omit $F_2$ &&\omit
$F_3$ &&\omit $F_4$ &&\omit  $F_5$ & \cr \noalign{\hrule} & $10$ &&
$SL(2)$ && $SO(2)$ &&   && $\bf 2$ && && $\bf 1$ &\cr & $8$ &&
$SL(3)\times SL(2)$ && $SO(3)\times SO(2)$ && $({\bf {\bar 3}},{\bf
2})$ &&$({\bf {\bar 3}},{\bf 1})$ &&$({\bf {1}},{\bf 2})$ && &\cr &
$7$ && $SL(5)$ && $SO(5)$ && ${\bf \bar {10}}$ && ${\bf {5}}$ &&
${\bf \bar {5}}$ && &\cr & $6$ && $SO(5,5)$ && $SO(5)\times SO(5)$
&& ${\bf {16}}$ && ${\bf { 10}}$  && && &\cr& $5$ && $E_6$ &&
$USP(8)$ && ${\bf { 27}}$   &&  && && &\cr & $4$ && $E_7$ && $SU(8)$
&& ${\bf {56}}$ && && && &\cr & $3$ && $E_8$ && $SO(16)$ && && && &&
&\cr \noalign{\hrule height 0.6pt} }}
\bigskip
\centerline{Table 1: $E_{n+1}$ , $I(E_{n+1})$ and representation of
the Field Strengths}
\bigskip

The scalars, denoted $\xi_E$, in $d$ dimensions belong to a non-linear realisation of $E_{n+1}$ with local
subgroup $I(E_{n+1})$ where $I(G)$ denotes  the Cartan involution invariant subgroup of $G$. These local
subgroups are given in Table 1.  Following the discussion of non-linear realisations given in appendix A we
find the transformations of   equation (A.6). Given a group element $g_E(\xi_E)$ of  $E_{n+1}$ we can use the
local transformation $I(E_{n+1})$ to cast it in the form
$$
g_E(\xi_E)= e^{\sum_{\vec \alpha
 >0}E_{\vec\alpha} \chi_{\vec\alpha} }e^{-{1\over \sqrt
2}\vec\phi\cdot \vec H} \eqno(2.28)
$$
where $E_{\vec\alpha} $ are the positive root and ${\vec H}$ the Cartan subalgebra  generators  of $E_{n+1}$.
The fields $\vec \phi$ and $\chi_{\vec\alpha}$ are the scalar fields of the theory which we have denoted
collectively  by $\xi_E$. The dynamics of the scalars are constructed, as usual,  out of the  Cartan form
$g_E^{-1}d g_E=P_E+Q_E $, where $Q_E$ lies in the Lie-algebra of $I(E_{n+1})$.

The  gauge fields transforms as linear representations of $E_{n+1}$;
their representations are given in Table 1. Note that care must be
taken for $d/2$-form field strengths as these generally only fill
out $E_{n+1}$ representations if their electromagnetic duals are
also included. However, it is desirable to use the scalar fields
$\xi_E$ to convert the fields strengths $F_E$ which belong to linear
realisations of $E_{n+1}$, into tensors denoted $\cal {F}_E$ which
transform non-linearly under $E_{n+1}$, using equation (A.8). We may
write the relation in the generic form
$$
|{\cal {F}}_E >= L(g^{-1}_E(\xi_E)) |F_E
>\ . \eqno(2.29)
$$
Under an $E_{n+1}$ transformation these change  as
$$
|{\cal F}_E >\to L(h^{-1}) |{\cal {F}}_E > \ ,\eqno(2.30)
$$
where $h\in I(E_{n+1})$. We can write $g_E(\xi_E)= g_{sl_2 \otimes
sl_n}(\xi_{sl_2 \otimes sl_n})g' $ where $g'$ contains the Cartan
and positive root generators of $E_{n+1}$ which are outside
$SL(2)\otimes SL(n)$. Therefore we can write
$$\eqalign{
|{\cal {F}}_E>&= L(g^{-1}_E(\xi_E)) |F >\cr &= L((g')^{-1}
L(g^{-1}_{sl_2 \otimes sl_n})(\xi_{sl_2 \otimes sl_n}) |F > \cr
&=\sum_{(\mu,\underline \lambda)} L((g')^{-1}) |{\cal F}^{(\mu,
\underline \lambda)}_{sl_2 \otimes sl_n} > \cr\ .} \eqno(2.31)
$$
Hence the $E_{n+1}$ non-linear realisations  ${\cal {F}}_E$ that
appear in the $E_{n+1}$ formulation of the theory can be written as
$L((g')^{-1})$ acting on the non-linear realisations ${\cal
F}^{(\mu,\underline \lambda)}_{sl_2 \otimes sl_n}$. The superscript
${(\mu,\underline \lambda)}$ are the highest weights of the
different $SL(2)\otimes SL(n)$ representations that arise in the
decomposition of the linear representation $F$, that is
$F=\sum_{(\mu,\underline \lambda)}  { F}^{(\mu,\underline
\lambda)}_{sl_2 \otimes sl_n} $.

We will primarily be interested in the scalar fields associated with the Cartan subalgebra of $E_{n+1}$. The
subalgebra $SL(n)\otimes SL(2) $ has $n$ such fields   $\underline \phi$ and $\phi$ which are associated with
the nodes $1\ldots ,n-1$ and node $n+1$ of the $E_{n+1}$ Dynkin diagram respectively. The remaining Cartan
field in $E_{n+1}$ is $\rho$ and this is associated with the deleted noded, that is the node $n$. Restricting
$g$ to the Cartan sub-algebra, denoted $g_c=e^{-{1\over \sqrt 2}\vec \phi \cdot \vec H}$ we find that
$$\eqalign{
L(g_c^{-1}) |\vec \lambda  >&= e^{+{1\over \sqrt 2}\vec \lambda
\cdot \vec \phi} |\vec \lambda  >\cr &=e^{{1\over \sqrt 2}(\vec
\lambda )_n\rho} e^{+{1\over \sqrt 2}\underline \lambda \cdot
\underline \phi} e^{+{1\over \sqrt 2}(\vec \lambda )_{n+1}
\phi}|\vec \lambda> \cr & =e^{{1\over \sqrt 2} (\vec \lambda
)_n\rho}L(g_{sl_2\otimes sl_n c} ^{-1})  |\vec \lambda  >\ ,}
\eqno(2.32)
$$
when acting on a state in a representation of $E_{n+1}$ with weight $\vec \lambda$. Here $(\vec \lambda )_n$ is
the nth component of $\vec \lambda $. We are interested in comparing the $E_{n+1}$ formulation of the higher
derivative corrections in $d$ dimensions with those obtained by dimensional reduction from ten dimensions, both
of which can be written in terms of non-linear realisation of $SL(2)\otimes SL(n)$ symmetry, {\it i.e.} in
terms of ${\cal F}^{(\mu,\underline \lambda)}_{sl_2\otimes sl_n}$. Consequently, it is the difference which is
of most interest, namely the $e^{-{1\over \sqrt 2} (\vec \lambda )_n\rho}$ factors. In the $E_{n+1}$
formulation these arise by decomposing the  $E_{n+1}$  building blocks ${\cal {F}}$ as in equation  (2.31) and
then using equation (2.32) while in the dimensional reduction they arise from the metric ansatz of equation
(2.1).

We assume that the  higher derivative corrections to the IIB theory are   invariant under a  discrete $E_{n+1}$
symmetry. The fields transform in the same way as for the IIB supergravity theory in $d$ dimensions, but under
the discrete group. The terms in the $d$ dimensional effective action will be of the generic form
$$
\int d^d x\det e \partial^{l_0} R^{{l_R\over 2}} P^{l_1}_E({\cal {F}}_{E \mu_1})^{l_1} ({\cal {F}}_{E \mu_1
\mu_2})^{l_2}\Phi_E\ldots \ , \eqno(2.33)
$$
where $ {\cal {F}}_{E \mu_1},\ldots $ are the $E_{n+1}$ non-linear realisations constructed in equation  (2.31)
and $\Phi_E$ is  function of the scalars $\xi$ which transforms under the discrete symmetry as
$$
\Phi_E\to D(h^{-1}) \Phi_E\ , \eqno(2.34)
$$
for  $h\in I(E_{n+1})$ and $D(h)$ being in the representation that $\Phi_E$ belongs to. However, $\Phi_E$  has
a non-holomorphic dependence  on the scalars and we will refer to it as a non-holomorphic automorphic form.

A formulation of automorphic forms which transform as in equation (2.34) was given in reference [29]. To
construct such a non-holomorphic automorphic form for a discrete group $G$ one chooses a linear representation
of $G$ denoted $|\psi>$ and considers $|\varphi>=L(g^{-1})|\psi>$ where $g(\xi)$ is an element of $G$ that is
subject to the transformations of equation (A.2), that is it is a non-linear realisation and $|\varphi>$ is the
non-linear realisation constructed from $|\psi>$ using equation (A.8). The automorphic form is a suitable
function of $\varphi$. The simplest case is that of a scalar automorphic form that is given by
$$
\Phi= \sum_{|\psi>\ne 0} {1\over <\varphi|\varphi>^s}\ . \eqno(2.35)
$$

For our case $G=E_{n+1}$ and $\xi$ are the scalar fields of the
theory, include those associated with Cartan subalgebra which we
have labeled by $\vec \phi = (\phi,\rho,\underline\phi)$. To leading
order the automorphic form will have a dependence on these scalars
which we denote by
$$
\Phi_E\sim e^{-\sqrt 2 \vec \lambda_\Phi\cdot \vec \phi}\ , \eqno(2.36)
$$
where $ \vec \lambda_\Phi$ a  weight of the representation. For the automorphic form of equation (2.35)
$\Phi\sim e^{- \sqrt 2 s\vec \lambda_H\cdot \vec \phi}$ where $\vec \lambda_H$ is the highest weight of the
represenation used to build the automorphic form.

 In this paper we will want to compare the
terms in the effective action of equation (2.33) in their $E_{n+1}$ formulation with those obtained from the
dimensional reduction of the higher derivative terms in ten dimensions given in equation (2.19). This will
allow us to place restrictions on the
   automorphic form $\Phi_E$ in $d$ dimensions and in particular the
weights $ \lambda^\phi$ that can appear in it.  For almost all terms
this will require the  decomposition of the $E_{n+1}$
representations that occur into $SL(2)\otimes SL(n)$
representations.

The simplest examples  are terms in the effective action of equation (2.33) that only involve powers of the the
Riemann tensor in $d$ dimensions since the Riemann tensor is a singlet of $E_{n+1}$. This contribution comes
from the dimensional reduction of the similar term in ten dimensions, namely that of equation (2.19) with only
$\hat l_R=l_R$ non vanishing. Since the Riemann tensor, in tangent frame, possess two powers of the inverse
vierbein we find a factor of $e^{-2\alpha \rho}$ for each Riemann tensor and a factor of $e^{2\alpha \rho}$
from $\det \hat e$. From the automorphic form in ten dimensions we find, at leading order, a factor of
$e^{-{\hat s} \phi}$. Thus from dimensional reduction we find in $d$ dimensions the term
$$
\int d^d x \det e R^{{l_R\over 2}} \Phi_E e^{-{\hat s} \phi-(l_R-2)\alpha\rho}\ . \eqno(2.37)
$$
Comparing this with the
$E_{n+1}$ formulation in  $d$ dimensions which is of the form $ \int
d^d x \det e R^{{l_R\over 2}}\hat \Phi_E $ we see that the
additional factor of $\phi$ and $\rho$ must arise from the
automorphic form $\Phi_E $ and so we find that,
$$
\vec \lambda_\Phi= \left({\hat s\over \sqrt 2}, \alpha {(l_R-2)\over \sqrt 2}, \underline 0\right)\ .
\eqno(2.38)
$$
From equation (2.21) we have $\hat s= {1\over 4}(l_R-2-8g)$ and taking the leading contribution at $g=0$ we
conclude that $ \vec \lambda_\Phi={1\over 4}(l_R-2)\vec \lambda^{n+1}$ where $\lambda^{n+1}= ({1\over \sqrt 2},
{1\over 2x},\underline 0)$ and we have used the  relation $x^{-1}=4\sqrt 2\alpha$. Thus the automorphic form
has the leading order behaviour $\Phi\sim e^{-\sqrt 2 {1\over 4}(l_R-2)\vec \lambda^{n+1}\cdot \vec \phi}$. Hence for  terms which contain only the Riemann curvature it is  
straight forward to to compute the leading behaviour of the automorphic form. In what follows we will  
carry out this calculation for all possible terms, but as we will see  
this involves some much more sophisticated group theory.

In order to study the remaining terms. We consider the possible building blocks that arise in the dimensional
reduction from ten dimensions and compare these with those in the $E_{n+1}$ formulation. As we have explained
above the latter can be expressed in terms of non-linear realisations of $SL(2)\otimes SL(n)$ which agree with
the same objects found from dimensional reduction. The difference arises from the $\rho$ dependence. To find
this difference we must decompose the  representations of $E_{n+1}$ into those of $SL(2)\otimes SL(n)$. We do
this following the techniques [41-43] developed for the study of the $E_{11}$ symmetry. As mentioned above,
deleting the node $n$ in the Dynkin diagram of $E_{n+1}$ results in the algebras $SL(2)\otimes SL(n)$. We may
write the simple roots of $E_{n+1}$ as
$$
\vec \alpha_{n+1}= (\beta_1, 0,0), \ \vec  \alpha_n=(0,x,\underline 0)-\vec \nu,\ \vec \alpha_i=
(0,0,\underline \alpha_i),\ i=1,\ldots , n-1 \ ,\eqno(2.39)
$$
where $\vec \nu= (\mu_1,0,\underline
0)+(0,0,\underline \lambda^{n-2})$. Also, the $\underline \alpha_i$
and $\underline\lambda^i$   are the simple roots  and fundamental
weights of $SL(n)$ and $\beta_1=\sqrt 2$ and $\mu={1\over \sqrt 2}$
the simple root and fundamental weight of $SL (2)$. Demanding that
$\vec \alpha_n^2=2$ we find that $x=\sqrt{{8-n\over 2n}}=(4\sqrt
2\alpha)^{-1}$.

The fundamental weights of $E_{n+1}$,   denoted $\lambda^a,
a=1,\ldots , n+1$, satisfy $\alpha_a\cdot \lambda^b=\delta_{a,b}$
and are given by
$$
\vec \lambda^i=(0,{1\over x}\underline \lambda^{n-2}\cdot \underline \lambda^i , \underline \lambda^i),\qquad
\vec \lambda^{n}=(0,{1\over x},\underline 0),\qquad  \vec \lambda^{n+1}=(\mu,{1\over 2 x},\underline 0)\ .
\eqno(2.40)
$$

Any root of $E_{n+1}$ can be written as
$$
\vec  \alpha=n_c \vec \alpha_n +m\vec \beta_1 +\sum_i n_i\vec \alpha_i= n_c (0,x,0) -\vec  \lambda \
,\eqno(2.41)
$$
where $ \vec \lambda= n_c\vec  \nu-\sum_i n_i (0,0,\underline
\alpha_i)-m (\beta_1,0,\underline 0)$. The latter is a weight of
$SL(2)\otimes SL(n)$. If a representation of $SL(2)\otimes SL(n)$
occurs in the decomposition of the adjoint representation of
$E_{n+1}$ its highest weight must occur as  one of the $\lambda$'s
for some positive integers $m$,  $n_i$ and $n_c$.  We refer to the
integer $n_c$ as the level and we can analyse the occurrence of
highest weights level by level using the techniques of references
[41-43].  Clearly, at level zero {\it i.e } $n_c=0$ we have just the
adjoint representation of $SL(2)\otimes SL(n)$. The result is that
the adjoint representation of $E_{n+1}$ contains the adjoint
representation of $SL(2)\otimes SL(n)$ at $n_c=0$ together with the
following highest weight representations of $SL(2)\otimes SL(n)$
$$
\matrix {n_c=1 & n_c=2 & n_c=3 & n_c=4 \cr (\mu,\underline \lambda^{2})&  (0,\underline \lambda^{4})&
(\mu,\underline\lambda^{6})&  (0,\underline \lambda^{1}+\underline\lambda^{n-7})&\cr} \ . \eqno(2.42)
$$

Thus the weights in the adjoint representation of $E_{n+1}$ then
have the from
$$\eqalign{
&([\beta_1], 0,\underline 0)\ ,   (0, 0,[\underline \alpha_1+\ldots +\underline \alpha_{n_1}])\ , \
([\mu_1],x,[\underline \lambda^{1}])\ ,\cr &  \   (0,2x,[\underline \lambda^{4}])\ ,\
([\mu_1],3x,[\underline\lambda^{6}])\ ,    (0,4x, [ \underline \lambda^{1}+\underline \lambda^{n-7}])\ .}
\eqno(2.43)
$$

These correspond to the adjoint  of $SL(2)\otimes SL(n)$ at $n_c=0$  as well as the generators
$$
\matrix {n_c=1 & n_c=2 & n_c=3 & n_c=4 \cr R^{\alpha ij} & R^{i_1\ldots i_4} &   R^{\alpha, i_1\ldots i_6} &
R^{j_1\ldots j_7 i}\cr}\ . \eqno(2.44)
$$
The maximum value of $n_c$ that contributes is
$n_c=1,2,2,3,4$ for $n=3,4,5,6,7$ respectively as is clear from the
index structures of the generators.  The reader may verify that once
the additional negative root generators are included  this
collection of generators has the correct count of generators for
$E_{n+1}$ for $n=3,\ldots , 7$.

The Cartan forms of $E_{n+1}$ belong to the adjoint representation and so using  equation (2.43) we find that
the coset  component $P_{Ea}$ decomposes into the Cartan forms $P_{sl_n\otimes sl_2 \mu}$ of $SL(2)\otimes
SL(n)$ at $n_c=0$ and
$$
\matrix {n_c=1 & n_c=2 & n_c=3 & n_c=4 \cr P_{sl_2\otimes sl_n \mu\alpha  ij} & P_{ sl_2\otimes sl_n \mu
i_1\ldots i_4} & P_{\mu \alpha , i_1\ldots i_6} & P_{sl_2\otimes sl_n \mu j_1\ldots j_7 i}\cr} \ .\eqno(2.45)
$$
The Cartan form contains the factor $e^{{1\over \sqrt 2}\vec\phi\cdot\vec\alpha}$, contained in the $g^{-1} $
part of $ g^{-1}\partial_\mu g$, and so using equation (2.43)  we find that the level $n_c$ contribution comes
with the factor
$$
e^{{1\over \sqrt 2}n_c x\rho}= e^{2\alpha \rho n_c {8-n\over n}} \ .\eqno(2.46)
$$

The ten dimensional origin of the first two terms of equation (2.44) is obvious given their index structure and
they are  contained in the blocks ${\cal F}_{sl_2\otimes sl_n \alpha \mu ij}$ and $F_{sl_2\otimes sl_n \mu
i_1\ldots i_4}$ respectively that come from the dimensional reduction of the three form and five form field
strengths respectively. The fourth term of equation (2.44) only occurs for $d=3$ and $d=4$ and in these
dimensions it arises as the dual of the three from field strength, more precisely the dual of $ {\cal
F}_{sl_2\otimes sl_n\  \alpha \nu_1\nu_2 \nu_3}$ and $ {\cal F}_{sl_2\otimes sl_n\  \alpha \nu_1\nu_2 i}$
respectively. Alternatively, one can think of the fourth term  as arising from the dimensional reduction of the
field strength ${\cal F}_{ sl2\alpha \nu_1\dots \nu_7}$. The final term in equation (2.44) only occurs in $d=3$
dimensions, that is for $E_8$, and it arises as  the dual  of the graviphoton $\partial_{[a} h_{b]}{}^i$.  At
the end of section three we showed that calculating  the powers $e^\rho$ from the original field, or its dual,
gave the same result. As such we will calculate it from the Cartan forms  of equation (2.45). We observe that
these carry one $d$ dimensional spacetime index and $2n_c$ internal indices and according to the discussion
around  equation (2.24) we find a factor of
$$
e^{-\rho(\alpha+2n_c\beta )}=e^{2\alpha \rho n_c {8-n\over
n}}e^{-\alpha\rho} \ ,\eqno(2.47)
$$
for each contribution.

Thus for each factor of the Cartan form $P_{E\mu}$ in the $d$ dimensional effective action we find  an
additional factor of $e^{-\alpha\rho}$ in the dimensionally reduced action compared to the $E_{n+1}$
formulation. This result, taken together with the previous result for factors of the Riemann tensor, is
consistent with the rule that   for each spacetime derivative in $d$ dimensions we get an additional factor of
$e^{-\alpha\rho}$.

To treat the other building blocks in the same way  we must learn how to decompose more general representations
of $E_{n+1}$ into those of $SL(2)\otimes SL(n)$.  To do this we use the technique of reference [44,45]. If one
wants to consider the fundamental representation $\vec \lambda^i$ of $E_{n+1}$ associated with the node labeled
$i$ we add a new node, denoted $\star$, to the $E_{n+1}$ Dynkin diagram which is connected to the node labeled
$i$ by a single line to construct the Dynkin diagram for an enlarged algebra of rank $n+2$. Deleting the
$\star$-node we recover the $E_{n+1}$ Dynkin diagram and the $\vec\lambda^i$ of $E_{n+1}$ is found in the
adjoint representation of the  enlarged algebra provided we keep only contributions at level $n_\star=1$. Thus
we find the decomposition of the fundamental representation of $E_{n+1}$ into representations of $SL(2)\otimes
SL(n)$ by decomposing the adjoint representation of the enlarged algebra but deleting the additional node and
keeping only contributions with $n_\star=1$ and deleting node $n$ but keeping all levels of $n_c$. The level
one states are a representation as the commutator preserves the level and so the commutator of the level zero
generators , that is the adjoint representation of $E_{n+1}$, with the level one states give again level one
states. It is the desired representation since the lowest state contains $\underline \lambda^i$. For the
details see reference [44,45].

The weights of the $\vec \lambda^i$ representation of $E_{n+1}$ can
be written in the  form
$$
\left([\mu], n_c x-{1\over
x}\underline\lambda^{n-2}\cdot\underline\lambda^i ,[\underline
\lambda ]\right)\ , \eqno(2.48)
$$
except for $i=n$ for which it is of the form $([\mu], n_c x-{1\over x} ,[\underline \lambda ])$ Here
$(\mu,\underline\lambda )$ is the highest weight of the $SL(2)\otimes SL(n)$ representation that occurs. We
note that $ \vec\nu\cdot\vec\lambda^i= {2i\over n}$ for $i\le n-2$, $ \vec\nu\cdot\vec\lambda^{n-1}=
{(n-2)\over n}$ and $ \vec\nu\cdot\vec\lambda^{n+1}={1\over 2}$.

Next  we will treat  the two form field strengths in the $d$
dimensional effective action in a similar way. The one form gauge
field, from which they are constructed, belong to the $\vec
\lambda^1$ representation of $E_{n+1}$. The $\vec \lambda^1$
representation of $E_{n+1}$ decomposes into $SL(2)\otimes SL(n)$
representations as follows
$$\eqalign{
&\matrix {n_c=0&n_c=1 & n_c=2 & n_c=3 & n_c=4&n_c=4&n_c=4 \cr
(0,\underline \lambda^{1}) &(\mu,\underline \lambda^{n-1})&
(0,\underline \lambda^{n-3})& (\mu,\underline\lambda^{n-5})&
(2\mu,\underline \lambda^{n-7}) &(\mu,\underline
\lambda^{n-7})&(0,\underline \lambda^{n-1}+\underline
\lambda^{n-6})\cr}\cr & \matrix {n_c=5&n_c=6 & n_c=7 &n_c=8\cr
(\mu,\underline \lambda^{n-2}+\underline \lambda^{n-7})&
(0,\underline \lambda^{n-4}+\underline \lambda^{n-7})&
(\mu,\underline \lambda^{n-6}+\underline \lambda^{n-7})&
(0,\underline \lambda^{n-1}+2\underline \lambda^{n-7})\cr}\ .}
\eqno(2.49)
$$
The reader may verify that one finds the correct dimensions of the $\vec \lambda^1$ representation,  that is
16, 27,  56 and 248 for $n=4,5,6 $ and $7$. The weights of the $\vec\lambda^1$ representation are given by
$$
(0, {2\over nx},\underline [\lambda^1])\ , ([\mu_1] , {2\over
nx}-x,[\underline \lambda^{n-1}])\ , (0, {2\over
nx}-2x,[\underline\lambda^{n-3}]) \ ,([\mu_1] , {2\over
nx}-3x,[\underline \lambda^{n-5}])\ ,
$$
$$
(2[\mu_1], {1\over 2x}-4x, [\underline\lambda^{n-7}]) \ , (0,
{2\over nx}-4x, [\underline\lambda^{n-7}]) \ , (0, {2\over
nx}-4x,[\underline\lambda^{1}]+[\underline\lambda^{n-6}])\ ,
$$
$$
([\mu_1], {2\over
nx}-5x,[\underline\lambda^{n-2}+\underline\lambda^{n-7}]) \ , (0,
{2\over nx}-6x,[\underline\lambda^{n-4}+\underline\lambda^{n-7}])\ ,
$$
$$
([\mu_1], {2\over
nx}-7x,[\underline\lambda^{n-6}+\underline\lambda^{n-7}]) \ ,
([\mu_1], {2\over
nx}-8x,[\underline\lambda^{1}+2\underline\lambda^{n-7}]) \eqno(2.50)
$$

These correspond to two form  field strengths take the form
$$
\matrix {n_c=0&n_c=1 & n_c=2 & n_c=3 & n_c=4&n_c=4&n_c=4 \cr {\cal F}_{\mu_1a_2  }^i& {\cal F}_{\alpha \mu_1a_2
i }&  {\cal F}_{\mu_1a_2 i_1i_2i_3}& {\cal F}_{\alpha \mu_1a_2 i_1\ldots i_5 }& {\cal F}_{\mu_1a_2 j,i_1\ldots
i_6 }&{\cal F}_{\mu_1a_2 (\alpha\beta),i_1\ldots i_7  }&{\cal F}_{\mu_1a_2 i_1\ldots i_7 }\cr}\ , \eqno(2.51)
$$
as well as higher level contributions. Since a two
form field strength is dual to a one form field strength in three
dimensions we only study two form field strengths in dimensions four
and above. This corresponds to $n\le6$ and so of the above field
strengths we only need those at levels  $n_c=3$ and the first term
in the above equation at level $n_c=4$.

We recognise the two form field strengths of equation (2.51) as the
dimensional reduction of   the metric, {\it  i.e.} the  graviphoton,
the three form, the five form for the first three entries. The
fourth entry arises from  the dual of the three form in $d=4$ and
$d=5$ while the only required level four field strength  is the dual
of the graviphoton.

Decomposing the rank two field strength in their $E_{n+1}$ representation, using   equations (2.29) and (2.32),
we find the factor
$$
e^{{1\over \sqrt 2}({\vec \nu\cdot \vec \lambda^1\over x}-n_c
x)\rho} =e^{{2\sqrt 2 \alpha\over n}(-4+n_c(8-n))\rho}\ ,
\eqno(2.52)
$$
for each rank two field strength at level $n_c$. We observe that the
above field strengths have two $d$-dimensional spacetime indices and
$2n_c-1$ internal indices and so the factor of $e^\rho$ that appears
when carrying out the dimensional reduction from ten dimensions is
$$
e^{-\rho(2\alpha+(2n_c-1)\beta )}= e^{{2\sqrt 2 \alpha\over
n}(-4+n_c(8-n))\rho} e^{-\alpha\rho}\ . \eqno(2.53)
$$
Evaluating this and comparing with the factor in equation (2.52) we find an additional factor of
$e^{-\alpha\rho}$ for each rank two field strength.

We now carry out the same analysis for the  rank three field strengths. We need only consider these field
strengths in dimensions $d\ge 6$, since in a lower dimension a  rank three field strength is dual to a lower
rank field strength. This is equivalent to $n\le 4$. The rank three field strength belong to the $\vec
\lambda^{n+1}$ representation of $E_{n+1}$. One finds that the weights in the $\vec\lambda^{n+1} $
representation of $E_{n+1}$ have the form
$$
([\mu_1], {1\over 2x},\underline 0)\ , (0, {1\over 2x}-x,[\underline\lambda^{n-2}])\ , ([\mu_1], {1\over
2x}-2x,[\underline\lambda^{n-7}])\ ,
$$
$$
(0, {1\over 2x}-3x,[\underline\lambda^{n-1}]+[\underline\lambda^{n-5}])\ , (0, {1\over
2x}-3x,[\underline\lambda^{n-6}])\ , ([\beta_1], {1\over 2x}-3x,[\underline\lambda^{n-6}])\ ,
$$
$$
([\mu_1], {1\over 2x}-4x,[\underline\lambda^{n-1}+\underline\lambda^{n-7}])\ , ([\mu_1], {1\over
2x}-4x,[\underline\lambda^{n-6}+\underline\lambda^{n-2}])\ ,
$$
$$
(0, {1\over
2x}-5x,[\underline\lambda^{n-4}+\underline\lambda^{n-6}])\ , (0,
{1\over 2x}-5x,[\underline\lambda^{n-1}+\underline\lambda^{n-2}
+\underline\lambda^{n-7}])\ ,
$$
$$
([\beta_1], {1\over
2x}-5x,[\underline\lambda^{n-3}+\underline\lambda^{n-7}])\ ,
$$
$$
([\mu_1], {1\over
2x}-6x,[\underline\lambda^{n-5}+\underline\lambda^{n-7}]) \ ,
([\mu_1], {1\over 2x}-6x,[\underline 2\lambda^{n-6}])\ ,
$$
$$
([\mu_1], {1\over
2x}-6x,[\underline\lambda^{n-1}+\underline\lambda^{n-4}
+\underline\lambda^{n-4}])\ ,\ldots .\eqno(2.54)
$$
The reader may like to verify that one has the correct count of
states for the $5$,$10$,$27$, and $133$-dimensional representations
of $SL(5)$, $SO(5,5)$, $E_6$ and $E_7$ respectively. For the first
few entries many contributions vanish as one has too many
antisymmetrised indices. To find the 3875 dimensional representation
of $E_8$ one must go further in the analysis.

The factor of $e^\rho$ associated with the term at $n_c$ is
$$
e^{-{1\over \sqrt 2}({\vec \nu\cdot \vec \lambda^{n+1}\over x}-n_c x) \rho} =e^{-{2\alpha\over
n}(n-2n_c(8-n))\rho}\ . \eqno(2.55)
$$
The corresponding field strengths carry three $d$
dimensional spacetime indices and $2n_c$ internal indices  and so we
find in the dimensionally reduced theory a factor of
$$
e^{-\rho(3\alpha+(2n_c)\beta )}=e^{-{\alpha\rho\over
n}(3n-2n_c(8-n))}\ . \eqno(2.56)
$$
Consequently for every rank three
field strength we find an additional factor of $e^{-\alpha\rho}$ in
the dimensionally reduced theory. The same conclusion holds for the
rank four field strengths.

Since one finds the same additional factor no matter what field strength one considers the above can be
summarised as follows,  for every derivative we find an  additional factor of $e^{-\alpha\rho}$ in the
dimensionally reduced theory. One also finds in the dimensionally reduced theory a $e^{-\hat s\phi}$ coming
from the ten dimensional $SL(2)$ automorphic form. Consequently, the excess in the dimensionally reduced theory
compared to that found in the $E_{n+1}$ formulation of equation (2.55), but  not taking into account  the
contribution of the $E_{n+1}$ automorphic form in $d$ dimensions in the latter formulation, is given by
$$
e^{-(l_T-2)\alpha\rho -\hat s \phi}\ , \eqno(2.57)
$$
where $l_T=\hat l_R+\hat l_1+\hat l_1'+\hat l_3^++\hat l_3^-+\hat
l_5$. The $-2$ part arises from the $\det \hat e$. This excess can
only come from the $E_{n+1}$ automorphic form. Demanding that all
the weights arising from dimensional reduction of the ten
dimensional theory appear in the $E_{n+1}$ formulation in $d$
dimensions we conclude that
$$
\vec \lambda_\Phi= \left({\hat s\over \sqrt 2}, \alpha {(l_T-2)\over
\sqrt 2}, \underline 0\right) =\left({l_T-2\over
4}+(l_{RR}-2g)\right)\vec \lambda^{n+1}+ \left({2g-l_{RR}\over
2}\right)\vec\lambda^n \ ,\eqno(2.58)
$$
where $l_{RR} =\hat l_1'+\hat l_3'+\hat l_5$ counts the number of RR
fields.

Let us consider higher derivative terms constructed only out of NS-NS fields, so that $l_{RR}=0$. Suppose also
that we look at terms which have a tree level, $g=0$,  contribution in ten-dimensions. In this case we find the
automorphic form in $d$ dimensions has the leading order behaviour $\Phi_E\sim e^{-\sqrt 2{(l_T-2)\over 4}\vec
\lambda^{n+1}}$. This strongly suggests that it is built from the $E_{n+1}$ representation with highest weight
$\vec \lambda^{n+1}$. This is the representation that the string charges of the $d$ dimensional theory belong
to.

The $SL(2,{\bf Z})$  Eisenstein automorphic form in ten dimensions contains two perturbative terms with dilaton
dependence $e^{-s\phi}$ and $e^{(s-1)\phi}$. If the first term possesses a value of $s$ that leads to a tree
level contribution then the second term leads to a genus $g={s-1/2}$ contribution. Above we considered the
effect of dimensionally reducing the tree level contribution, but one can also consider the second
contribution. One finds, substituting $g=s-1/2$ into (2.58), that the weight vector is
$$
\vec\lambda_\Phi=(1-s)\vec\lambda^{n+1} +
(s-1/2)\vec\lambda^n=s\vec\lambda^{n+1} - (s-1/2)\vec\alpha_{n+1}\ .
\eqno(2.59)
$$
However the first two  terms in the
perturbative contribution of the Eisenstein-like $E_{n+1}$
automorphic form in $d$ dimensions constructed using the
$\vec\lambda^{n+1}$ representation are of the generic form [30]
$$
\Phi_E \sim E_1e^{- \sqrt{2}s\lambda^{n+1}}+E_2e^{-
\sqrt{2}(s\vec\lambda^{n+1}-(s-1/2)\vec\alpha_{n+1})}\ , \eqno(2.60)
$$
where $E_1$ and $E_2$ are constants. It is pleasing to see that the
second term of the automorphic form in ten dimensions leads to the
correct second term in the $E_{n+1}$ automorphic form in $d$
dimensions.

We note that a similar calculation for dimensional reduction of type IIA string theory on an $n$-torus leads to
the same results as the the type IIB reduction considered here [46].

\bigskip
{\llarge 3. M-Theory}
\bigskip

Let us now perform a similar analysis for the dimensional reduction of higher derivative terms of M-theory.
Note that to compare with the previous section one must make the substitution $n\to n+1$. In addition the
values of $\alpha$ and $x$ in this section are different to those of section 2.

The Bosonic field content of M-theory consists of the graviton with
curvature $\hat R$ and a three form gauge field $\hat A_{\hat\mu
\hat\nu \hat\rho}$ out of which the four form field strength $\hat
F_{\hat \mu \hat \nu \hat \rho \hat \sigma}$ is constructed. At
lowest order in derivatives the low energy effective action may be
written
$$
\int d^{11} x \det \hat e \left( \hat R - {1\over 2\cdot 4!}\hat F^2
+\ldots  \right)\ .\eqno(3.1)
$$
where the ellipsis denote Fermion terms as well as a Chern-Simons-type term for $\hat A$. A generic higher
derivative correction in the $d=11$ low energy effective action of M-theory may be written,
$$
\int d^{11} x \det \hat e (\partial)^{\hat l_0} \hat R^{{\hat l_1\over 2}} \hat F^{\hat l_4}.\eqno(3.2)
$$
M-theory, dimensionally reduced on an $n$-torus, possesses an
$E_{n}$ symmetry in $d=11-n$ dimensions and shares the same manifest
$SL(n)$ symmetry through the non-linearly realised field strengths
and the Cartan forms in $d$ dimensions as the type IIB theory.
However, no dilatonic scalar is present in $d=11$ dimensions. Upon
dimensional reduction, a higher derivative term will pick up a
dependence on the $n$ diagonal components of the metric on the
$n$-torus $\rho$ and $\underline \phi$. We observe that the higher
derivative terms in the dimensionally reduced formulation carry a
factor of $e^{\sqrt{2} \vec w . \vec \phi}$ where the $n$ vectors
$\vec w$ and $\vec \phi$ recording the dilatonic scalar field
content and their associated weights are defined as
$$\eqalign{
\vec \phi &= \left( \rho, \underline \phi \right),\cr \vec w &=
\left( \kappa, \underline w \right).}\eqno(3.3)
$$
The general term in the $E_{n}$ formulation in $d$ dimensions is a polynomial in the non-linearly realised
field strengths $\cal F$, Cartan forms $P$ and curvature $R$ multiplied by an automorphic form $\Phi_E$
constructed out of some representation of $E_{n}$
$$ \int d^d x{\rm det}e\partial^{l_0}R^{{l_R\over
2}} P^{l_1}_{E\mu_1}({\cal {F}}_{E \mu_1 \mu_2})^{l_2}({\cal {F}}_{E \mu_1\mu_2\mu_3})^{l_3}
\Phi_E\ldots.\eqno(3.4)
$$
We will again determine the representation out of which the $E_{n}$ automorphic form is constructed in $d$
dimensions by comparing the dimensionally reduced formulation, with manifest $SL(n)$ symmetry, to that of the
$E_{n}$ formulation.  The Dynkin diagram for M-theory is (note that here we use a different labeling for the
nodes and hence the roots and weights are also labeled differently than in section 2)
$$
\matrix{ & & & & & & \bullet & \vec \alpha_n& & & \cr & & & & & & | & & & & \cr \bullet &  - &  \ldots & -
&\bullet&-&\bullet&-&\bullet&-&\bullet  \cr \vec \alpha_{1}& & & & \vec\alpha_{n-4} & &\vec \alpha_{n-3} & &
\vec\alpha_{n-2} && \vec \alpha_{n-1} \cr}
$$
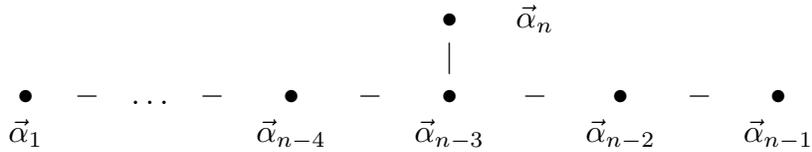
\bigskip \centerline{Fig. 3 Dynkin diagram for $E_{n+1}$ in M-theory labelling} \bigskip
The simple roots of $E_{n}$ may be written as
$$
\vec \alpha_{i} = \left(0, \underline\alpha_{i} \right), \ \ \ i =
1,...,n-1, \qquad \vec\alpha_{n} =\left(x, \underline 0 \right) -
\vec \mu,\eqno(3.5)
$$
where $\vec \mu = \left(0,\underline\lambda^{n-3} \right)$.  The variable $x$ associated with the $\rho$ factor
of the deleted node $n$ is evaluated via the inner products between the simple roots of $E_{n}$, given by the
corresponding Cartan matrix, one finds
$$
x=\sqrt{{9-n\over n}}=\left(3\sqrt{2} \alpha \right)^{-1}\ . \eqno(3.6)
$$
The fundamental weights of $E_{n}$, dual to the simple roots $\underline \alpha_{i}$, are
$$\eqalign{
\vec \lambda^{i} &= \left( {1\over x} \underline
\lambda^{i}\cdot\underline  \lambda^{n-3} , \underline \lambda^{i}
\right),  \cr \vec \lambda^{n} &= \left(  {1\over x}, \underline  0
\right).}\eqno(3.7)
$$
One may write any root of $E_{n}$ as
$$
\vec\alpha=n_{c} \vec \alpha_{n}  + \sum_{i=1}^{n-1} n_{i} \vec \alpha_{i} = n_{c}\left(x, \underline 0 \right)
- \vec \lambda\ ,\eqno(3.8)
$$
where $\vec \lambda = n_{c} \vec \nu  - \sum_{i=1}^{n-1} n_{i} \vec  \alpha_{i}$. As in the IIB theory, if a
representation of $SL(n)$ is present at some level $n_{c}$ in the adjoint representation of $E_{n}$, then its
highest weight may be written as $\vec \lambda$ for some combination of the integers $n_{c}$ and $n_{i}$. Level
$n_c=0$ contains the adjoint representation of $SL(n)$. The highest weight representations of $SL(n)$ at higher
levels are
$$
\matrix{
 n_c=1  & n_c=2 &  n_c=3 \cr
\underline \lambda^{ 3} &  \underline \lambda^{ 6} & \underline
\lambda^{ 1}+ \underline  \lambda^{n-8}  \cr }\ .\eqno(3.9)
$$
So the weights in the lower levels of the adjoint representation of $E_{n}$ are
$$
\left( 0, \left[\underline\alpha_{1} + ... + \underline\alpha_{n-1}
\right] \right) \ , \left( x, \left[ \underline  \lambda^{3} \right]
\right) \ , \left(  2 x, \left[ \underline  \lambda^6 \right]
\right) \ , \left( 3 x, \left[ \underline  \lambda^{ 1}  +
\underline  \lambda^{n-8} \right] \right)\ .\eqno(3.10)
$$
The decomposition of the Cartan form $P_{E}$, at a given level $n_{c}$ is found by examining the weights.  At
level $n_{c}=0$ the Cartan form $P_{E}$ contains the Cartan form of $SL(n)$ at higher levels the Cartan form
$P_{E}$ decomposes as follows
$$
\matrix{
 n_c=1 & n_c=2 & n_c=3  \cr
 P_{sl_ni_{1}i_{2} i_{3}}&  P_{sl_ni_{1}...i_{6}} & P_{sl_nj,i_{1}...i_{8}} \ .  \cr
} \eqno(3.11)
$$
The Cartan form contains the factor $e^{{1\over\sqrt{2}} \vec \phi . \vec \alpha}$, so at level $n_{c}$ we find
a factor of
$$
e^{{1\over\sqrt{2}}\left(  \left( n_c x \right)\rho  \right)} =
e^{\left( 3n_c\right)\alpha \rho \left( {9-n\over
n}\right)}.\eqno(3.12)
$$
With the natural ordering on the levels, we find the maximum level
that contributes is $n_{c}=1$ for $n=3,4$, $n_{c}=2$ for $n=5,6$ and
$n_{c}=3$ for $n=7$.  The $SL(n)$ Cartan forms $P_{SL(n)}$ originate
from the four form field strength $\hat G_{\mu\overline i_1
\overline i_2 \overline i_3}$ at level $n_{c}=1$, the dual of the
four form field strength at level $n_c=2$ and the graviphoton at
level $n_c=3$. These Cartan forms of $SL(n)$, arising upon
dimensional reduction, carry one $d$ dimensional spacetime index and
$\left( 3 n_c \right)$ internal indices.  Therefore, each Cartan
form of $SL(n)$, at level $n_{c}$, is multiplied by the factor
$$
e^{-\rho\left(\alpha +\left( 3n_c \right)\beta  \right)}=e^{\left(
3n_c \right) \alpha \rho \left({9-n\over n} \right)}e^{-\alpha
\rho}. \eqno(3.13)
$$
The two form field strengths lie in the representation of $E_{n}$
with highest weight $\vec \lambda^{1}$. Decomposing the $\vec
\lambda^{1}$ of $E_{n}$ into representations of $SL(n)$ level by
level, we find
$$
\matrix{
 n_c=0 & n_c=1 & n_c=2 & n_c=3 \cr
 \underline  \lambda^{1} &   \underline \lambda^{n-2}&
 \underline \lambda^{n-5} &   \underline  \lambda^{n-1} + \underline  \lambda^{n-7} . \cr
}\eqno(3.14)
$$
Therefore, for $n \leq 7$, the weights of the $\vec \lambda^{1}$
representation are
$$
\left(  {3\over nx}  , \left[\underline  \lambda^{1} \right] \right)
\ , \left( {3\over nx}   - x, \left[ \underline  \lambda^{n-2}
\right] \right) \ , \left( {3\over nx} - 2 x, \left[ \underline
\lambda^{n-5} \right] \right) \ , \left( {3\over nx} - 3  x, \left[
\underline \lambda^{n-1} + \underline \lambda^{n-7} \right] \right)
.\eqno(3.15)
$$
From the weights, we see that the corresponding two form field strengths, at each level, are
$$
\matrix{
 n_c=0 & n_c=1 & n_c=2 & n_c=3 \cr
{\cal F}^i_{\mu_1 \mu_2}&{\cal F}_{\mu_{1} \mu_{2} i_{1} i_{2}}  &
{\cal F}_{\mu_{1} \mu_{2} i_{1}...i_{5} } & {\cal F}_{\mu_{1}
\mu_{2} i , j_{1}...j_{7}} . \cr }\eqno(3.16)
$$
The two form field strengths appear in $d\geq 4$ dimensions.  In
$d=11-n$ dimensions one finds that all two form field strengths,
with associated level $n_c$, satisfying the constraint $n\leq 3n_{c}
-1$ will be present. We see that the two form field strength at
level $n_c=0$ arises through the dimensional reduction of the metric
and four form at levels $n_c=0$, $n_{c}=1$ respectively.  The two
remaining levels in the decomposition of the $\vec \lambda^{1}$ are
associated with the duals of the four form field strength and the
graviphoton at $n_{c}=2$ and $n_{c}=3$ respectively. Since the two
form field strengths in the $E_{n}$ lie in some representation of
$SL(n)$ at level $n_{c}$ in the decomposition of $\vec \lambda^{1}$
they carry a multiplicative factor of
$$
e^{{1\over\sqrt{2}}\left( \left(-{3\over nx} + n_c x  \right)\rho
\right)} = e^{-{9\over n} - n_c \alpha \rho \left({9-n\over
n}\right)}.\eqno(3.17)
$$
If we compare the multiplicative factor found through the
decomposition of the $\vec \lambda^{1}$ in the $E_{n}$ formulation
to the corresponding factor arising in the dimensionally reduced
formulation, where the two form field strengths carry two $d$
dimensional indices and $3n_c -1$ internal indices, and so appear
multiplied by the factor
$$
e^{-\rho\left(2 \alpha + \left( 3n_{c} -1 \right) \beta \right)} =
e^{-\alpha \rho} e^{-{9\over n} - n_c \alpha \rho \left({9-n\over
n}\right)},\eqno(3.18)
$$
we find that the two form field strengths in the dimensionally
reduced M-theory formulation carry a surplus factor of $e^{-\alpha
\rho}$. In the $E_{n}$ formulation the three form field strengths
lie in the representation with highest weight $\vec \lambda^{n-1}$.
One finds that the $\vec \lambda^{n+1}$ representation of $E_{n}$
decomposes, in the following way for $n\le  5$
$$
\matrix{
 n_c=0 & n_c=1 \cr
 \underline \lambda^{n-1} &   \underline \lambda^{n-4}  .\cr
}\eqno(3.19)
$$
We observe that, for $n \leq 5$, the weights in the $\vec \lambda^{n-1}$ representation of $E_{n}$ are
$$
\left( \left({n-3\over nx}\right) , \ \left[ \underline
\lambda^{n-1} \right] \right) \ , \left( {\left(n-3\right)\over nx}
- x ,  \left[ \underline \lambda^{n-4} \right] \right) .\eqno(3.20)
$$
The three form field strengths, at level $n_c$, are
$$
\matrix{ n_c=0 & n_c=1 \cr {\cal F}_{\mu_1 \mu_2 \mu_3 i}& {\cal
F}_{\mu_{1} \mu_{2} \mu_{3} i_{1}...i_{4}}. \cr}\eqno(3.21)
$$
The three form field strength occurring in the decomposition of the
$\lambda^{n-1}$ at level $n_{c}=0$ arises from the dimensional
reduction of the four form field strength, while the other, at level
$n_{c}=1$ is associated with the dual of the dimensionally reduced
four form field strength.  The three form field strengths at levels
$n_c=0,1$ appear in $d = 6,7$ dimensions, in $d=8$ only the $n_c=0$
three form field strength is present. The decomposition of the $\vec
\lambda^{n-1} $ of $E_{n}$, at level $n_c$, is multiplied by a
factor of
$$
e^{-{1\over\sqrt{2}}\left( {\left(n-3\right)\over nx}  - n_c x
\right)\rho } = e^{\left( \left(-3+{9\over n} \right) +n_c
\left({9-n\over n}\right)\right)\alpha \rho}.\eqno(3.22)
$$
The three form field strengths in the dimensionally reduced
formulation come with three spacetime indices and
$\left(3n_c+1\right)$ internal indices, therefore they carry a
factor of
$$
e^{-\rho\left(3 \alpha + \left(3n_{c}+1  \right) \beta  \right)} =
e^{-\alpha \rho} e^{\left( \left(-3+{9\over n}  \right) +n_c
\left({9-n\over n}\right)\right)\alpha \rho}.\eqno(3.23)
$$
In $d=11-n$ dimensions, the Cartan forms, field strengths and
curvatures lying in the $E_n$ representation may be constructed out
of the dimensionally reduced Cartan forms, field strengths and
curvatures with manifest $SL(n)$ symmetry.  For example, the two
form field strengths in $d=7$ dimensions lie in the $\bf { {10}}$ of
$E_{4}$, which may be constructed out of the two form field
strengths arising from dimensional reduction to $d=7$. Namely, the
graviphotons lying in the $\bf {4}$ and the dimensionally reduced
four form field strength $\hat G_{\mu_{1}\mu_{2}\overline i_1
\overline i_2 }$ lying in the $\bf {\underline  6}$ of $SL(4)$.
However, each of the dimensionally reduced terms carry an additional
factor of $e^{-\alpha \rho}$. Therefore, any product of Cartan
forms, field strengths and curvatures, in the $E_{n}$ formulation,
reconstructed using the appropriate dimensionally reduced terms,
will be multiplied by a surplus factor of
$$
e^{ - \left( l_{T}-2 \right) \alpha \rho  },\eqno(3.24)
$$
where $l_{T}$ is the total number of derivatives in the product.
This factor must be attributed to the automorphic form in the
$E_{n}$ formulation.  To leading order, we may write the automorphic
form in the $E_{n}$ formulation as $\Phi_{E_n} \sim e^{-\sqrt 2\vec
\lambda_\Phi \cdot \vec \phi}$.  Thus, one finds
$$
\vec \lambda_{\Phi} = \left( \alpha \left( {l^{T} - 2
\over\sqrt{2}}\right) , \underline  0  \right) = \left({l_{T} -
2\over 6} \right) \vec \lambda^{n}.\eqno(3.25)
$$

\bigskip

{\llarge 4. Discussion}
\bigskip

In this paper we have dimensionally reduced the higher derivative
terms of ten dimensional IIB theory and deduced the weight vectors
that are associated with the Cartan subalgebra fields of the
$E_{n+1}$ symmetry. Most of these weights are accounted for  once
the $d$-dimensional theory is expressed in terms of $E_{n+1}$
covariant building blocks involving the  Riemann tensor, field
strengths and derivatives of the scalars.  However, we also found
that there was always a remaining weight. This implies that
polynomials constructed only out of the field strengths are not
consistent with U-duality in the lower dimension. On the other hand
these additional weights can be accounted for in the $d$ dimensional
theory if they are attributed to an $E_{n+1}$ automorphic form. In
this way we obtained constraints on the automorphic forms that occur
in $d$-dimensions.

Carrying out this procedure we have found that the dimensional
reduction of the IIB higher derivative corrections implies that such
terms in $d$ dimensions should contain an automorphic   form
involving the  weight $\vec\lambda^{n+1}$, using the labeling of the
Dynkin diagram of Figure 1.  It is natural to think of this as the
highest weight of the representation used to construct the
automorphic form. This applies to all terms in a given dimension,
although  this does not mean that the same automorphic appears for
all terms. For terms that only contain the Riemann tensor and scalars the leading order weight can be readily deduced by counting the number of inverse metrics required, however for more general terms we needed  to perform a detailed group theory analysis. 

As the constraints we find  arise from considering the ten
dimensional theory we are in effect  considering terms that survive
the decompactification from $d$ dimensions, that is $\rho\to
-\infty$. We have focused particularly on the terms that arise at
tree level in ten dimensions. However we also saw that the
next-to-leading order contribution in ten-dimensions correctly
matched that of the $d$-dimensional automorphic form if the
$\vec\lambda^{n+1}$ representation is used for the case of
Einstein-like automorphic forms.

This result is in agreement with the results [17-36] found so far
for terms with  low numbers of spacetime derivatives in that  the
automorphic forms studied for these terms are constructed from the
$\vec\lambda^{n+1}$ representation. It is also  natural in that  the
string charges belong to the $\vec\lambda^{n+1}$ multiplet and the
discrete $E_{n+1}$ group acts naturally on these objects.

We also  performed a similar calculation from the viewpoint of
eleven-dimensional M-theory. We found that the automorphic forms
should contain the weight $\vec\lambda^{n-1}$, using the type IIB
labeling of the $E_{n+1}$ Dynkin diagram of Figure 1.  This is also
natural as membrane charges belong to the $\vec\lambda^{n-1}$
representation. It would be interesting to reconcile this result
with that from the IIB perspective. The automorphic forms contain
combinations of weights and one would have to find the combination
of weights predicted from the M-theory viewpoint in the automorphic
from constructed from the representation with highest weight
$\vec\lambda^{n+1}$ that it used in the type IIB theory. In this way
the M-theory analysis places a non-trivial constraint on the
automorphic forms.

A more radical possibility is that  the automorphic forms for these
different representations and suitable $s$ are actually the same. In
fact for $SL(5)$ the two representations are the $\bf 5$ of the
string and the $\bf \bar 5$ of the membrane,  the highest weight of
the former being minus the lowest weight of the latter, lead to
automorphic forms that are indeed related for suitable values of
$s$. Indeed, for the higher rank groups some correspondences of this
type for the automorphic forms corresponding to terms with  low
numbers of spacetime derivatives have already  have been conjectured
in [27,36]. This would require very considerable conspiracies since
the representations involved are quite different (including vastly
different dimensions).

Another possibility is that  a given higher derivative term can
involve more than one automorphic form based on different
representations. This has already been found to occur in seven
dimensions for the $\partial^4 R^4$ term [35].

\bigskip
{\llarge Acknowledgments}
\bigskip

Peter West would like to thank The Erwin Schroedinger International
Institute for Mathematical Physics and the Theoretical Physics
Department of the Vienna University of technology for their kind
hospitality in October 2009 when some of the work in this paper was
carried out.  This work has been supported by an STFC Rolling grant
ST/G000395/1.

\bigskip
{\llarge Appendix A: Non-linear Realisations}
\bigskip

In this appendix we  review of the  construction of non-linear
realisations in a form suitable to that used in this paper. We
consider a group $G$ with Lie algebra $Lie(G)$. $Lie(G)$ can be
split into the Cartan subalgebra with elements $\vec H$, positive
root generators $E_{\vec\alpha}$ and negative root generators
$E_{-\vec\alpha}$ with $\vec\alpha>0$. There exists a natural
involution, known as the Cartan involution, defined by
$$ \tau
:(\vec H,E_{\vec\alpha})\to -(\vec H, E_{-\vec\alpha})\ . \eqno
(A.1)
$$
To construct the  non-linear realisation we must specify a
subgroup $H$ (not to be confused with  the generators of the Cartan
subgroup which are denoted by $\vec H$). For us this is defined to
be  the subgroup left invariant under the Cartan involution, i.e.
$H=\{g\ \in \ G:\tau(g)=g \}$. In terms of the Lie algebra $Lie(H)$
it is all elements $A$ such that $A=\tau(A)$.

The non-linear realisation is constructed from group elements $g(x) \in G$ that depend on spacetime that are
subject to the transformations
$$
g(x)\to g_0 g(x) h^{-1}(x)\ , \eqno (A.2)
$$
where $g_0\in G$ is constant and $h(x)\in H$ depends on spacetime. We may write the group element in the form
$$
g(x) = e^{\sum_{\vec\alpha>0} \chi_{\vec\alpha} E_{\vec\alpha}}
e^{-{1\over \sqrt{2}}\vec\phi\cdot\vec H} e^{\sum_ {\vec\alpha>0}
u_{\vec\alpha} E_{-\vec\alpha}}\ ,\eqno(A.4)
$$
 but using the local transformation we can bring it
to the form
$$
g(\xi) = e^{\sum_{\vec\alpha>0} \chi_{\vec\alpha} E_{\vec\alpha}}
e^{-{1\over \sqrt{2}}\vec\phi\cdot\vec H}\ .\eqno(A.5)
$$
Here we use  $\xi = (\vec\phi,\chi_{\vec\alpha})$ as a generic
symbol for all the scalar fields, which are functions of spacetime,
that parameterize the coset representative. Under a rigid $g_0\in G$
transformation $g(\xi) \to g_0 g(\xi)$ this form for the coset
representative is not preserved. However one can make a compensating
transformation $h(g_0,\xi)\in H$ that returns $g_0g(\xi) $ into the
form of equation (A.5);
$$
g_0 g(\xi)h^{-1}(g_0,\xi) = g(g_0\cdot \xi)\ .\eqno (A.6)
$$
This induces a non-linear action of  the group $G$ on
the scalars; $\xi \to g_0\cdot \xi$.

We will also need   a  linear representation of $G$.  Let
$\vec\mu^i$, $i=1,...,N$ be the weights   of the representation and
$|\vec\mu^i>$ be a corresponding  states. We choose $\vec\mu^1$ to
be  the highest weight and so the corresponding state satisfies
$E_{\vec\alpha}|\vec\mu^1>=0$ for all simple roots $\vec\alpha$.
The states in the rest of the representation are polynomials of
$F_{\vec \alpha}= E_{-\alpha}$ acting on the highest weight state.

We consider states of the form  $|\psi>=\sum_i\psi_i |\vec\mu^i>$.
Under the action $U(g_0)$ of the group $G$ we have
$$
|\psi>\to U(g_0)|\psi>=L( g_0^{-1}) \sum_i \psi_i|\vec\mu^i> \equiv (U (g_0)\psi_i) |\vec\mu^i>= \sum_{i,j}
D_i{}^j (g_0^{-1})\psi_j |\vec\mu^i> \ ,\eqno (A.7)
$$
where $L(g_0) $ is the expression of the group element $g_0$ in terms of the Lie algebra elements which now act
on the states of the representation in the usual way. We note that the action of the group on the components
$\psi_i$ is given by $ \psi_i\to U(g_0)\psi_i= \sum_j D_i{}^j (g_0^ {-1})\psi_j $ which is the result expected
for a passive action. The advantage of using the states to discuss the representation is that  we can use the
action of the Lie algebra elements $L(g_0) $ on the states to compute the matrix $ D_i{}^j$ of the
representation and  deduce properties of the representation in general.

Given any linear realisation, such as the one in equation  (A.8), we
can construct a non-linear realisation by
$$
|\varphi(\xi)> = \sum \varphi_i |\vec\mu^i> =L(g^{-1}(\xi))|\psi> =
e^{\sum_{\vec\alpha>0} e^{{1\over \sqrt{2}}\vec\phi\cdot\vec
H}}e^{-\sum_{\vec\alpha>0}\chi_{\vec \alpha} E_{\vec\alpha}} |\psi>
\ ,\eqno (A.8)
$$
where $g(\xi)$ is the group element of the  non-linear realisation
in equation (A.5). Under a group transformation $U(g_0)$ it
transforms as
$$\eqalign{
U(g_0)|\varphi(\xi)> &=  L(g^{-1}(\xi))U(g_0)|\psi> =L(g^{-1}(\xi)) L(g_0^{-1} ) |\psi> \cr &= L( (g_0
g^{-1}(\xi))|\psi>\cr &= L(h^{-1})|\varphi(g_0\cdot\xi)> \ ,}\eqno (A.9)
$$
using equation (A.2). In terms of the component fields we find that $ \varphi_i (\xi)= \sum_j
D_i{}^j(g^{-1}(\xi))\psi_j$ and $U(g_0)\varphi_i (\xi)= \sum_j D_i{}^j((h)^{-1})\varphi_j (g_0\cdot \xi)$. The
reader can find the example of $SL(2)$ worked out in section 2.1.

\bigskip
{\large {References}}
\bigskip

\item{[1]}
     I.~C.~G.~Campbell and P.~C.~West,
     %``N=2 D = 10 Nonchiral Supergravity And Its Spontaneous Compactification,''
     Nucl.\ Phys.\ B {\bf 243}, 112 (1984).
     %%CITATION = NUPHA,B243,112;%%

\item{[2]}
     F.~Giani and M.~Pernici,
     %``N=2 Supergravity In Ten-Dimensions,''
     Phys.\ Rev.\ D {\bf 30}, 325 (1984).
     %%CITATION = PHRVA,D30,325;%%

\item{[3]}  M.~Huq and M.~A.~Namazie,
     %``Kaluza-Klein Supergravity In Ten-Dimensions,''
     Class.\ Quant.\ Grav.\  {\bf 2}, 293 (1985)
     [Erratum-ibid.\  {\bf 2}, 597 (1985)].
     %%CITATION = CQGRD,2,293;%%

\item{[4]}
J.~H.~Schwarz and P.~C.~West,
     %``Symmetries And Transformations Of Chiral N=2 D = 10 %Supergravity,''
     Phys.\ Lett.\ B {\bf 126}, 301 (1983).
     %%CITATION = PHLTA,B126,301;%%

\item{[5]}
     P.~S.~Howe and P.~C.~West,
     %``The Complete N=2, D = 10 Supergravity,''
     Nucl.\ Phys.\ B {\bf 238}, 181 (1984).
     %%CITATION = NUPHA,B238,181;%%

\item{[6]}
     J.~H.~Schwarz,
     %``Covariant Field Equations Of Chiral N=2 D = 10 Supergravity,''
     Nucl.\ Phys.\ B {\bf 226}, 269 (1983).
     %%CITATION = NUPHA,B226,269;%%

\item{[7]}
     E.~Cremmer, B.~Julia and J.~Scherk,
     %``Supergravity Theory In 11 Dimensions,''
     Phys.\ Lett.\ B {\bf 76}, 409 (1978).
     %%CITATION = PHLTA,B76,409;%%

\item{[8]}
E.~Cremmer and B.~Julia,
%``The N=8 Supergravity Theory. 1. The Lagrangian,''
Phys.\ Lett.\ B {\bf 80}, 48 (1978).
%%CITATION = PHLTA,B80,48;%%

\item{[9]}
N.~Marcus and J.~H.~Schwarz,
     %``Three-Dimensional Supergravity Theories,''
     Nucl.\ Phys.\ B {\bf 228} (1983) 145.
     %%CITATION = NUPHA,B228,145;%%

\item{[10]}
B.~Julia and H.~Nicolai,
     %``Conformal internal symmetry of 2d sigma-models coupled to %gravity and  a
     %dilaton,''
     Nucl.\ Phys.\ B {\bf 482}, 431 (1996)
     [arXiv:hep-th/9608082].
     %%CITATION = HEP-TH 9608082;%%

\item{[11]}
B.~Julia, in {\it Vertex Operators and Mathematical Physics},
Publications of the Mathematical Sciences Research Institute no3.
Springer Verlag (1984); in {\it Superspace and Supergravity}, ed.
S.~W.~Hawking and M.~ROcek, Cambridge University Press (1981)

\item{[12]} C. Teitelboim, Phys. Lett. B167 (1986) 69.

\item{[13]}  R. Nepomechie , Phys. Rev. D31, (1984) 1921;

\item{[14]}A.~Sen,
     %``Electric magnetic duality in string theory,''
     Nucl.\ Phys.\ B {\bf 404}, 109 (1993)
     [arXiv:hep-th/9207053].
     %%CITATION = HEP-TH 9207053;%%

\item{[15]}
A.~Font, L.~E.~Ibanez, D.~Lust and F.~Quevedo,
     %``Strong - weak coupling duality and nonperturbative effects in string theory,''
     Phys.\ Lett.\ B {\bf 249}, 35 (1990).
     %%CITATION = PHLTA,B249,35;%%

\item{[16]}
C.~M.~Hull and P.~K.~Townsend,
     %``Unity of superstring dualities,''
     Nucl.\ Phys.\ B {\bf 438}, 109 (1995)
     [arXiv:hep-th/9410167].
     %%CITATION = HEP-TH 9410167;%%

\item{[17]}
M.~B.~Green and M.~Gutperle,
     %``Effects of D-instantons,''
     Nucl.\ Phys.\ B {\bf 498}, 195 (1997)
     [arXiv:hep-th/9701093].
     %%CITATION = HEP-TH 9701093;%%

\item{[18]}
     M.~B.~Green, M.~Gutperle and P.~Vanhove,
     %{\sl One loop in eleven dimensions},
     Phys.\ Lett.\ B {\bf 409} (1997) 177
     [arXiv:hep-th/9706175].
     %%CITATION = HEP-TH 9706175;%%

\item{[19]}M.~B.~Green and S.~Sethi,
     %``Supersymmetry constraints on type IIB supergravity,''
     Phys.\ Rev.\ D {\bf 59}, 046006 (1999)
     [arXiv:hep-th/9808061].
     %%CITATION = HEP-TH 9808061;%%

\item{[20]} M.~B.~Green, H.~h.~Kwon and P.~Vanhove,
     %``Two loops in eleven dimensions,''
     Phys.\ Rev.\ D {\bf 61}, 104010 (2000)
     [arXiv:hep-th/9910055].
     %%CITATION = HEP-TH 9910055;%%

\item{[21]}M.~B.~Green and P.~Vanhove,
     %``Duality and higher derivative terms in M theory,''
     JHEP {\bf 0601}, 093 (2006)
     [arXiv:hep-th/0510027].
     %%CITATION = HEP-TH 0510027;%%

\item{[22]} M.~B.~Green, J.~G.~Russo and P.~Vanhove,
     %``Non-renormalisation conditions in type II string theory and maximal supergravity,''
     arXiv:hep-th/0610299.
     %%CITATION = HEP-TH 0610299;%%

\item{[23]}
A.~Basu,
     %``The D**10 R**4 term in type IIB string theory,''
     arXiv:hep-th/0610335.
     %%CITATION = HEP-TH 0610335;%%

\item{[24]}
N.~Berkovits and C.~Vafa,
     %``Type IIB R**4 H**(4g-4) conjectures,''
     Nucl.\ Phys.\ B {\bf 533}, 181 (1998)
     [arXiv:hep-th/9803145].
     %%CITATION = HEP-TH 9803145;%%

\item{[25]}
E.~Kiritsis and B.~Pioline,
     %``On R**4 threshold corrections in type IIB string theory and
(p,q) string instantons,''
     Nucl.\ Phys.\ B {\bf 508}, 509 (1997)
     [arXiv:hep-th/9707018].
     %%CITATION = HEP-TH 9707018;%%

\item{[26]}
  A.~Basu,
  %``The D^4 R^4 term in type IIB string theory on T^2 and U-duality,''
  Phys.\ Rev.\  D {\bf 77} (2008) 106003
  [arXiv:0708.2950 [hep-th]].
  %%CITATION = PHRVA,D77,106003;%%

\item{[27]}
   N.~A.~Obers and B.~Pioline,
   %``Eisenstein series and string thresholds,''
   Commun.\ Math.\ Phys.\  {\bf 209}, 275 (2000)
   [arXiv:hep-th/9903113].
   %%CITATION = CMPHA,209,275;%%

\item{[28]}
N.~Lambert and P.~West,
%``Enhanced coset symmetries and higher derivative corrections,''
Phys.\ Rev.\ D {\bf 74}, 065002 (2006) [arXiv:hep-th/0603255].
%%CITATION = HEP-TH 0603255;%%

\item{[29]}
N.~Lambert and P.~West,
%``Duality groups, automorphic forms and higher derivative corrections,''
Phys.\ Rev.\ D {\bf 75}, 066002 (2007) [arXiv:hep-th/0611318].
%%CITATION = HEP-TH 0611318;%%

\item{[30]}
 N.~Lambert and P.~West,
  %``Perturbation Theory From Automorphic Forms,''
  arXiv:1001.3284 [hep-th].
  %%CITATION = ARXIV:1001.3284;%%

\item{[31]}
  L.~Bao, M.~Cederwall and B.~E.~W.~Nilsson,
  %``Aspects of higher curvature terms and U-duality,''
  Class.\ Quant.\ Grav.\  {\bf 25} (2008) 095001
  [arXiv:0706.1183 [hep-th]].
  %%CITATION = CQGRD,25,095001;%%

\item{[32]}
L.~Bao, J.~Bielecki, M.~Cederwall, B.~E.~W.~Nilsson and D.~Persson,
  %``U-Duality and the Compactified Gauss-Bonnet Term,''
  JHEP {\bf 0807} (2008) 048
  [arXiv:0710.4907 [hep-th]].
  %%CITATION = JHEPA,0807,048;%%

\item{[33]}
L.~Bao, A.~Kleinschmidt, B.~E.~W.~Nilsson, D.~Persson and
B.~Pioline,
  %``Instanton Corrections to the Universal Hypermultiplet and Automorphic Forms
  %on SU(2,1),''
  arXiv:0909.4299 [hep-th].
  %%CITATION = ARXIV:0909.4299;%%

\item{[34]}
 B.~Pioline and D.~Persson,
  %``The automorphic NS5-brane,''
  arXiv:0902.3274 [hep-th].
  %%CITATION = ARXIV:0902.3274;%%

\item{[35]}
 M.~B.~Green, J.~G.~Russo and P.~Vanhove,
  %``Automorphic properties of low energy string amplitudes in various
  %dimensions,''
  arXiv:1001.2535 [hep-th].
  %%CITATION = ARXIV:1001.2535;%%

\item{[36]}  B.~Pioline,
  %``R**4 couplings and automorphic unipotent representations,''
  arXiv:1001.3647 [hep-th].
  %%CITATION = ARXIV:1001.3647;%%

\item{[37]}
N.~Berkovits,
  %``New higher-derivative R**4 theorems,''
  Phys.\ Rev.\ Lett.\  {\bf 98} (2007) 211601
  [arXiv:hep-th/0609006].
  %%CITATION = PRLTA,98,211601;%%

\item{[38]}  P.~C.~West,
     %``E(11) and M theory,''
     Class.\ Quant.\ Grav.\  {\bf 18}, 4443 (2001)
     [arXiv:hep-th/0104081].
     %%CITATION = HEP-TH 0104081;%%

\item{[39]}
I.~Schnakenburg and P.~C.~West,
   %``Kac-Moody symmetries of IIB supergravity,''
   Phys.\ Lett.\ B {\bf 517}, 421 (2001)
   [arXiv:hep-th/0107181].
   %%CITATION = HEP-TH 0107181;%%

\item{[40]}A.~Obers and B.~Pioline,
  %``U-duality and M-theory,''
  Phys.\ Rept.\  {\bf 318} (1999) 113
  [arXiv:hep-th/9809039].
  %%CITATION = PRPLC,318,113;%%

\item{[41]}
M.R Gaberdiel, D. I. Olive and P. West.
% ``A Class of Lorentzian Kac-Moody Algebras",
Nucl. Phys. {\bf B 645} (2002) 403-437, hep-th/0205068.

\item{[42]}  T.~Damour, M.~Henneaux and H.~Nicolai,
     %``E(10) and a 'small tension expansion' of M theory,''
     Phys.\ Rev.\ Lett.\  {\bf 89}, 221601 (2002)
     [arXiv:hep-th/0207267].
     %%CITATION = HEP-TH 0207267;%%

\item{[43]}
P. West,
%{\sl Very-extended $E_8$ and $A_8$ at low levels},
Class. Quant. Grav. {\bf 20} (2003) 2393,  hep-th/0307024.

\item {[44]}  A. Kleinschmidt and P. West,
%{\it  Representations of G+++ and the role of space-time},
JHEP 0402 (2004) 033,  hep-th/0312247.

\item{[45]} P. West,
%{\it $E_{11}$ origin of Brane charges and U-duality multiplets},
JHEP 0408 (2004) 052, hep-th/0406150.

\item{[46]} F. Gubay and P. West, to appear.

\end